\documentclass[useAMS,usenatbib]{mn2e}
\usepackage{natbib}
\usepackage{fixltx2e}
\usepackage{graphicx}
\usepackage{epstopdf} 
\usepackage{epsfig}
\usepackage{float}
\usepackage[fleqn]{amsmath}

\title[An in-depth study of PKS J0334-3900]{Using head-tail galaxies to constrain the intracluster
magnetic field: an in-depth study of PKS J0334-3900}
\author[Pratley et al.]
{Luke Pratley$^1$, Melanie Johnston-Hollitt$^1$, Siamak Dehghan$^1$ and Ming Sun$^2$ 
\\
$^1$ School of Chemical \& Physical Sciences, Victoria University of Wellington, PO Box 600, Wellington, 6140, New Zealand\\
$^2$ Eureka Scientific, Inc., 2452 Delmer Street, Suite 100, Oakland, CA 94602, USA
}
\date{March 2013}

\pagerange{\pageref{firstpage}--\pageref{lastpage}} \pubyear{2013}

\def\LaTeX{L\kern-.36em\raise.3ex\hbox{a}\kern-.15em 
    T\kern-.1667em\lower.7ex\hbox{E}\kern-.125emX}

\begin{document}

\label{firstpage}

\maketitle

\begin{abstract}
We present an in-depth, multi-wavelength study of the radio galaxy PKS J0334-3900, which resides at the centre of Abell 3135. The spectro-polarimetric radio observations are combined with spectroscopic optical and X-ray data to illustrate the use of Head-Tail radio galaxies in revealing properties of the intracluster medium (ICM). Australia Telescope Compact Array (ATCA) observations at 1.4, 2.5, 4.6 \& 8.6 GHz are presented with a detailed analysis of the morphology and spectral indices which give physical parameters to constrain the dynamical history of the galaxy. Using these constraints we produce a simulation of PKS J0334-3900. We find that this particular Head-Tail morphology can be induced via a combination of orbital motion due to a binary companion and the relative motion through the ICM. New Chandra images of A3135 are presented from which we obtain a cluster electron density of $n_{e,0} = (1.06 \pm 0.11) \times 10^{-3} \textrm{cm}^{-3}$, a global temperature of 2.4 $^{+0.51}_{-0.38}$ keV and a lower limit to the radio jet power of PKS J0334-3900 of 1.6 $\times 10^{44}$ erg/s. A new redshift analysis of the cluster from available spectroscopic data demonstrates A3135 to be comprised of galaxies with $0.058 \leq z < 0.066$ and gives a new mean cluster redshift of 0.06228 $\pm$ 0.00015. We also uncovered a background subgroup between $0.066 \leq z <  0.070$. Additionally, ATCA Stokes Q and U data of Abell 3135 were used to obtain rotation measure values along the line of sight to PKS J0334-3900.  Using our simulation we are able to infer the distance between the jets along the line of sight to be 154 $\pm$ 16 kpc, which when combined with the difference in rotation measure between the jets provides a novel new way to estimate the average magnetic field within a cluster. A lower limit to the cluster magnetic field was calculated to be 0.09 $\pm$ 0.03 $\mu$G.  From these results, we have shown that different techniques can be combined from observations of Head-Tail galaxies to infer information on the cluster environment, showing them to be an important class of objects in next generation all sky surveys.

\end{abstract}


\section{Introduction}
Abell 3135 (A3135) is a galaxy cluster located at the northern edge of the Horologium-Reticulum supercluster (HRS). It is a relatively unstudied system, with little work done to date on the cluster dynamics. The most distinguishing feature of Abell 3135 is the spectacular Head-Tail radio galaxy, PKS J0334-3900, found close to the cluster centre. Head-Tail (HT) galaxies are radio galaxies that have their jets bent in the same direction with the core elliptical galaxy and bent jets resembling a head and tail. The morphology and radio power of HT galaxies put them in the transition region between the Fanaroff-Riley classes I and II \citep{fanaroffriley}.

HT galaxies are found mostly within or on the periphery of galaxy clusters (though they are not only found in clusters, \citet{xizhen95}). They have been found in regions of galaxy clusters that are dense and turbulent \citep{mao09}, at a rate of about 1 - 2 per cluster in the local ($z<0.1$) Universe. In the more distant Universe, tailed radio galaxies are believed to act as signposts for over densities in large-scale structure, with HT galaxies being associated with galaxy clusters up to z $\sim$ 1 \citep{blanton03}. There is evidence that this association continues to the limits of both cluster and tailed galaxy detection, of  z $\sim$ 2 \citep{Dehghan11a}. 

Since the time of their discovery 40 years ago \citep{miley72} the morphology of HT galaxies has been believed to be the result of environmental interactions, commonly due to the surrounding intra-cluster medium (ICM) \citep{burns98}. In particular, the bent morphology is the result of ram pressure from the HT's relative motion with respect to the ICM. This ram pressure can arise via the in-fall into the cluster potential well \citep{GunnGott} or from winds associated with ``weather" in ICM which could arise either from mergers or mass accretion driven activity. We can therefore use the morphology of an HT to study the nearby cluster environment. 

In recent years, the use of tailed radio galaxies as environmental probes has gained momentum as a method for cluster detection \citep{blanton03,smolcic07,mao10}, examining the dynamics of individual clusters \citep{pfrommer11}, measuring the density \citep{freeland08} and velocity flows  \citep{douglass11} in the ICM, and for probing cluster magnetic fields \citep{clarke01,eilek02,johnston-hollitt03,vogt03,guidetti08}. 

However, to date, no study has combined the morphology and linear polarisation of an HT galaxy to improve estimates of cluster magnetic fields. We present a novel new method, where we combine knowledge about the morphology and linear polarisation of PKS J0334-3900 to probe the cluster magnetic field. From the morphology of PKS J0334-3900, a simple simulation is used to study how the morphology was generated by the environment. From this simulation, we can estimate the distance between the jets of PKS J0334-3900 along the line of sight. The distance and intervening ICM between the jets are used to explain the difference in average rotation measure between the jets. The net rotation measure between the jets provides an estimate of the cluster magnetic field.

 The paper is presented as follows: Section 2 presents a spectroscopic analysis of A3135. Section 3 presents the current knowledge of PKS J0334-3900; Section 4 contains details of the radio and X-ray observations and the results found in this study of PKS J0334-3900; Sections 5 presents the simulation of PKS J0334-3900, made using the morphology obtained from the results, this simulation is then used with the rotation measure values to estimate the cluster magnetic field. Section 6 presents the conclusion.

We adopt a standard set of cosmological parameters throughout with $H_0 = 73 kms^{-1}Mpc^{-1}$, $\Omega_m = 0.27$ \& ${\Omega}_{\Lambda}=0.73$. At the redshift of PKS J0334-3900, 0.062310 $\pm$ 0.000097 \citep{collins95}, 1 arcsecond is 1.185 kpc.

\section{Optical Properties of A3135}
Spectroscopic redshifts were collected from the literature to determine the cluster velocity distribution. From this distribution a new average redshift of the cluster was calculated, showing PKS J0334-3900 to be at the centre of this distribution.

\subsection{Spectroscopic Analysis of A3135}

\begin{figure}
\begin{center}
\vspace{-0.7cm}
\vbox{\includegraphics[width=9.25cm]{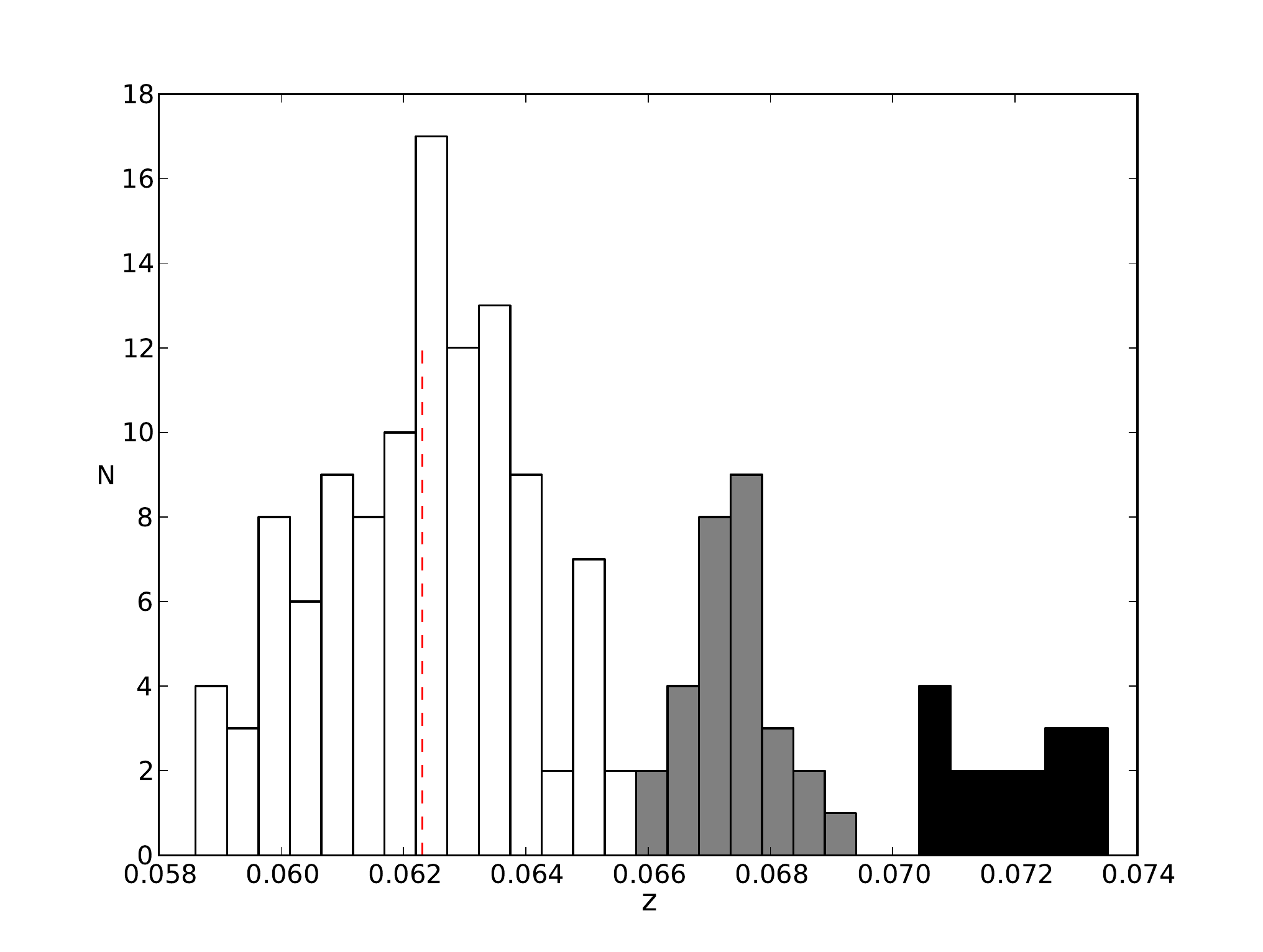}}{\vspace{-0.3cm}\includegraphics[width=9.25cm, height=7.7cm]{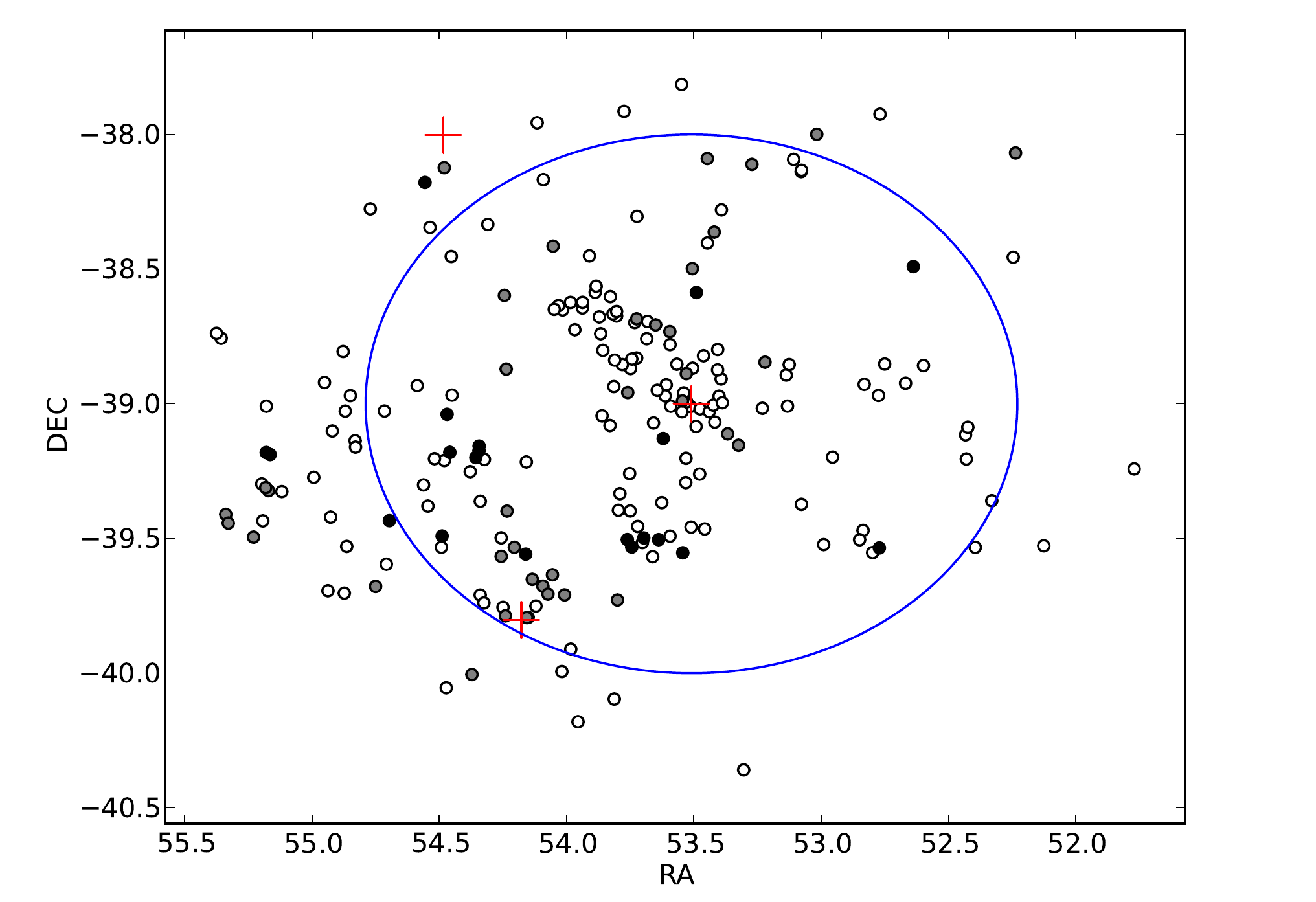}}
\caption{Top: Histogram of redshifts for the 155 galaxies within one degree of A3135 and with redshifts between 0.058 and 0.075. The plot has been split into three subgroups with the following colours; white is $0.058 \leq z<0.066$, grey is $0.066 \leq z<0.070$ and black is $0.070 \leq z<0.075$. The dashed red line gives the redshift of PKS J0334-3900 showing it is in the central bin of the main cluster component (shown as the white subgroup). Bottom: scatter plot of available redshifts surrounding A3135, colour coded as per the binning in the histogram. The circle denotes the galaxies within one degree of the cluster which are included in the histogram above. Crosses show the location A3135 and other clusters in the HRS.}\label{fig:RedShiftDist}
\end{center}
\end{figure}

There are 310 spectroscopic redshifts available in the literature \citep{galli93,collins95,loveday96,shectman96,dalton97,katgert98,ratcliffe98,tucker00,colless01,colless03,way05,jones09} within 1 degree of A3135, of these 155 have values between 0.058 and 0.075, which has been defined to be the extent of the ``Kinematic Core" of the HRS \citep{fleenor05}. Figure \ref{fig:RedShiftDist} shows the velocity histogram of the 155 galaxies known to be associated with A3135 \& the HRS (top panel) in addition to their respective position on the sky (bottom panel). The spectroscopic data were split into three components, denoted on Figure \ref{fig:RedShiftDist} via different colours. Most of the galaxies have redshifts between 0.058 and 0.066 (white bins) and mark the main part of A3135. The other two groups are located at $0.066 \leq z<0.070$ (grey bins), which represent galaxies in a nearby background group, and $0.070 \leq z<0.075$ (black bins) which are more distant galaxies separated by a significant velocity gap, and thus are not expected to be part of A3135. As shown on the figure, a large number of the galaxies with redshifts $0.058 \leq z<0.066$ are close to the centre of A3135, showing that the centre of A3135 is denser, as expected. The over density of galaxies in the south-east is due to the nearby cluster Abell A3142, centred at ${\rm RA_{J2000}}$ = 03:36:17.3, ${\rm DEC_{J2000}}$ = -40:38:09 (marked by a cross on the plot). Also visible in the top of the lower panel of Figure \ref{fig:RedShiftDist} is A3145, denoted by a cross. All three clusters are members of the HRS.

The most reliable published value of the redshift for A3135 is $0.0642$ found using 82 galaxies \citep{struble99}. However, we find using 110 galaxies with redshifts between $0.058 \leq z<0.066$ (the white histogram in \ref{fig:RedShiftDist}) that the average redshift is lower, with $z = 0.06228 \pm 0.00015$. The higher 
previously published value of the redshift can be explained by the inclusion of the previously unknown sub-group at $0.066 \leq z<0.070$ in the Struble calculation. This new mean redshift is in much better agreement with the redshift of the central Dominant galaxy, PKS J0334-3900, shown as the red dashed line in Figure \ref{fig:RedShiftDist}, falling within the peak cluster redshift bin. 

\section{PKS J0334-3900}
Several different radio telescopes have been used to observe PKS J0334-3900, at frequencies ranging from 80 MHz to 8.6 GHz. In particular, low resolution observations have been performed with the Very Large Array (VLA) at 1.4 GHz \citep{condon98} \& 4.89 GHz \citep{ekers89} and the Molonglo Observatory Synthesis Telescope (MOST) at 408 MHz \citep{schilizzi75} \& 843 MHz \citep{jones92}.  Previous radio observations have classified PKS J0334-3900 as a wide-angle tail (WAT) galaxy, with a power consistent with an FRI \citep{govoni00}. In this work, we present higher resolution observations from the Australia Telescope Compact Array (ATCA) at 1.4, 2.5, 4.8 \& 8.6 GHz.

PKS J0334-3900's optical host is a giant elliptical galaxy with an apparent Johnson blue magnitude of $b_J = 16.00 \pm 0.06$ \citep{hambly01b,hambly01a}
. It has been identified as the brightest cluster galaxy \citep{sun09}, with a redshift of 0.062310 $\pm$ 0.000097 \citep{collins95}. The optical host has a redshift slightly less than the previous cluster average of 0.0642 \citep{struble99}, but in good agreement with the new cluster redshift, 0.06228 $\pm$ 0.00015, derived here. 

PKS J0334-3900 is powered by a supermassive black hole with a mass of $3.5 \times 10^8$ solar masses \citep{bettoni03}, making it fairly typical for a radio galaxy \citep{woo02}. PKS J0334-3900 is at the centre of A3135, meaning it is in the most dense region of the cluster, this is consistent with HTs found in other clusters in the HRS \citep{mao09}.

\section{Observations and Results}
The results presented in this section were derived using the radio and X-ray observations of PKS J0334-3900 and the ICM respectively. The results derived from the radio \& X-ray observations include: the Stokes I images, the spectral index, polarisation angle maps, rotation measure values,  and a Chandra X-ray image of the ICM. Each of these results contains useful information for understanding how PKS J0334-3900 interacts with its environment; the Stokes I images display the morphology, the spectral index can be used to estimate the age of the source, the polarisation angle maps  help understand the morphology by showing the direction of flow within the jets, the rotation measures are required for understanding the cluster magnetic field, and the X-ray image is used to estimate the electron density. 

\subsection{Radio Observations}

\begin{figure*}
\includegraphics[width=180mm]{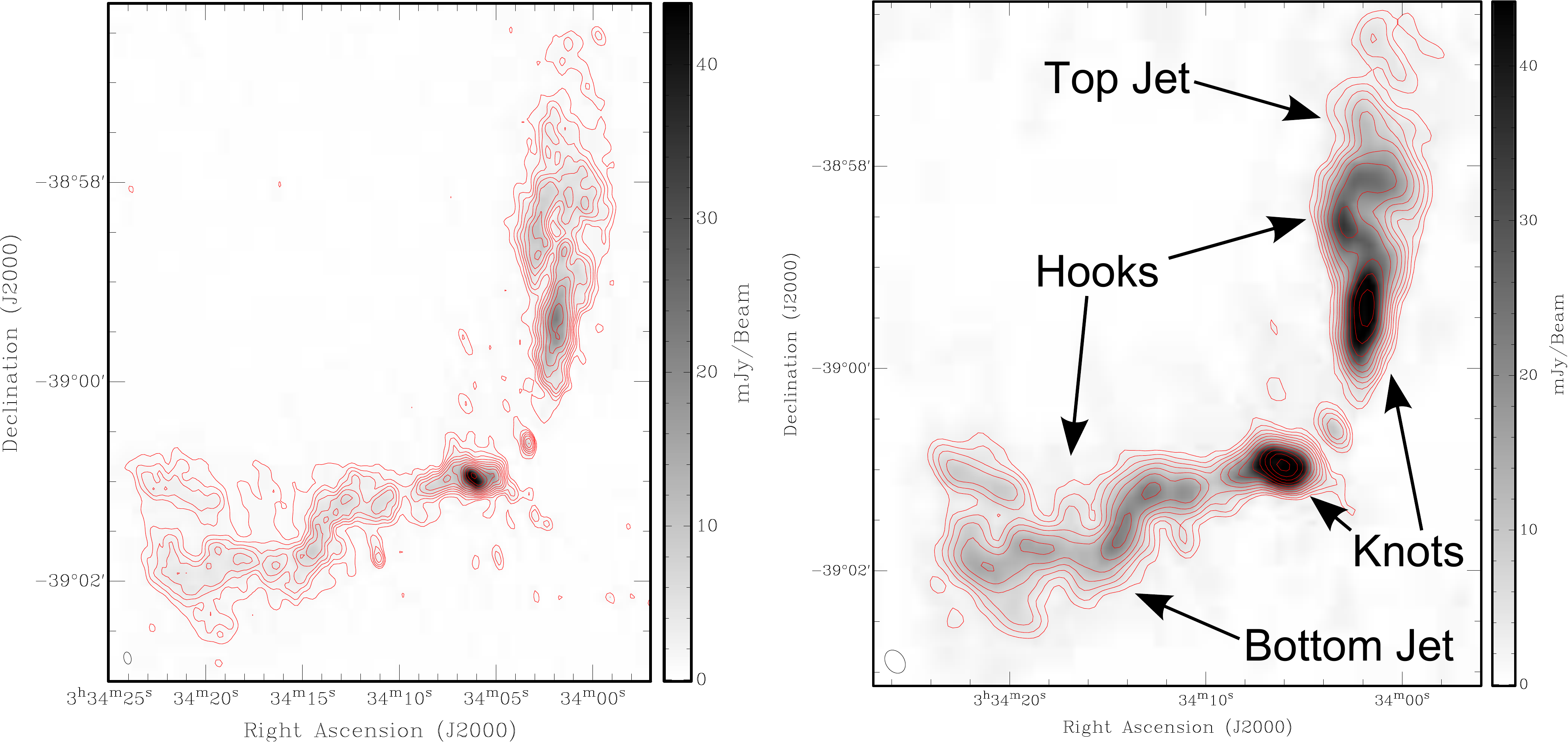}
\caption{Stokes-I 1384 MHz images of PKS J0334-3900. The image on the left is at full resolution (beam size $7.46^{\prime\prime} \times 4.41^{\prime\prime}$) and has an RMS noise of 0.37mJy beam$^{-1}$. The image on the right has had a taper applied to give a beam size of $14.8^{\prime\prime} \times 11.0^{\prime\prime}$ resulting in an RMS noise of 0.9 mJy  beam$^{-1}$. In both images the contours start at 3 times the RMS noise and increase in multiples of $\sqrt{2}$.}\label{fig:cont1384}
\end{figure*}

\begin{figure}
\includegraphics[width=7cm,angle=270]{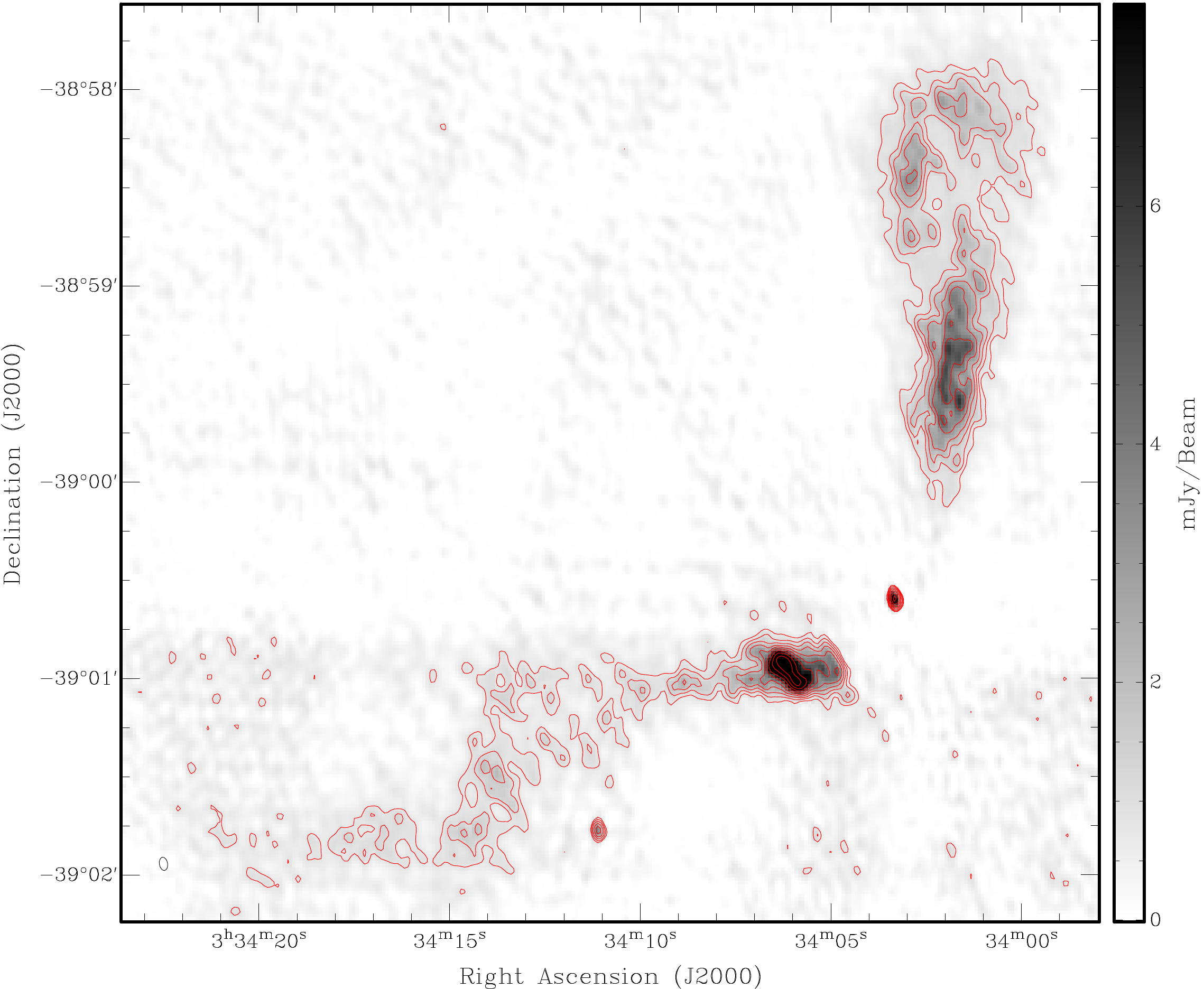}
\caption{Stokes-I data at 2496 MHz for PKS J0334-3900 with a beam size of $4.1^{\prime\prime} \times 2.5^{\prime\prime}$ and an RMS noise of  0.266 mJy beam$^{-1}$. The red contours show the shape of the jet from this Stokes-I image and increase in multiples of $\sqrt{2}$ of the RMS noise until the $6^{th}$ contour.}\label{fig:cont2496}
\end{figure}

\begin{table}
\caption{Details of the observations of PKS J0334-3900 using the ATCA. Column 1 is the array configuration; column 2 is the date; column 3 is the integration time; column 4 gives the central observing frequency of the 128 MHz ATCA band and column 5 is the resultant UV range listed as an interval.} \label{tab:ReductionInfo}
\begin{tabular}{ l l l l l }
\hline
Conf. &Date &Time &Freq. &UV range\\ 
& & (Min.)&(MHz) &(k$\lambda)$ \\
\hline
6A & 2001 May 20&59&  1384&  [0.5, 28.5] \\  
& & &2496&[1.8, 48.2] \\ 
& &31.2&4800&[4,96]\\
& & &8640&[6,172]\\ 
\hline
1.5A & 2001 Aug 3 \& 7& 76 &  1384 & [0.3, 21.1] \\ 
& & &2496 &[0.6,36.2] \\
& &63.1 &4800 &[0.8,23],[30,69.5] \\ 
& & &8640&[3,43],[55,128] \\
\hline
750A & 2001 Jul 13& 79.7 &1384& [0.1, 3.5],[9, 17.9]\\ 
& & &2496 &[0.2,6],[15.4,30.5] \\
& &55.1&4800&[0.5,12],[30.2,59.5] \\
& & & 8640& [3,21][55,105]\\
\hline
375 & 2001 Jun 15 & 75.4 & 1384 & [0, 2.1], [16.5, 28.5] \\ 
& & &2496&[0.1, 4],[28, 49] \\ 
& &64.5&4800&[0.4,8.5],[56,96] \\
& & &8640&[0.2,13],[100,173] \\
\hline
\end{tabular}
\end{table}

All ATCA observations of PKS J0334-3900 were extracted from the archive (PI - Saripalli). They were performed in 2001 using a phase centre of ${\rm RA_{J2000}}$ = 03:34:07.180, ${\rm DEC_{J2000}}$ = -39:00:03.190. The observations used the frequencies 1384, 2496, 4800 \& 8640 MHz, and a bandwidth of 128 MHz over 32 channels. The channels on the edge of the band were removed, leaving an effective bandwidth of 96 MHz. Due to cross-channel interference in the 1384 MHz observation, the 8 MHz interval centred at 1408 MHz was also removed.

 Each observation used four baseline configurations (see Table \ref{tab:ReductionInfo}), providing excellent uv-coverage. Taking all configurations into account, the total integration time was 290 minutes at 1384 \& 2496 MHz, and 214 minutes at 4800 \& 8640 MHz.

The flux densities were scaled according to observations of PKS B1934-638, and the assumed total flux densities at 1384, 2496, 4800 \& 8640 MHz are $14.94 \pm 0.01$, $11.14 \pm 0.01$, $5.829 \pm 0.005$ \& $2.841 \pm 0.006$ Jy, respectively \citep{reynolds94}. The time variations in complex antenna gains and bandpass were calibrated using observations of the unresolved source PKS B0332-403.

The observations were calibrated using MIRIAD, while following the standard ATCA pre-CABB reduction method \citep{sault95}.
The Stokes intensity images were also restored using MIRIAD. A summary of the resolution and RMS noise for the Stokes I images shown in Figures \ref{fig:cont1384}, \ref{fig:cont2496} and \ref{fig:StartOfJet} can be found in Table \ref{tab:ContInfo}.

\begin{table}
\caption{Details of continuum Stokes I images made and shown in Figures \ref{fig:cont1384}, \ref{fig:cont2496} and \ref{fig:StartOfJet} and also used for spectral index calculations, but not shown. At 1384 and 8640 MHz two images were made, one at full resolution and one after applying a taper in the uv-plane.}\label{tab:ContInfo}
\begin{tabular}{l l l l l}
\hline
Frequency &Synthesized Beam Size & PA & Stokes I $\sigma$\\ 
(MHz)&(arcseconds)&$\circ$ &(mJy b$^{-1}$)\\ 
\hline
1384& $7.459 \times  4.406$&11.40&$0.367$\\
    & $14.81 \times  10.99$ &0&$0.867$\\
\hline
2496& $4.088 \times 2.507$&9.826&$0.266$\\
\hline
4800& $1.799 \times 1.434$&26.32&$0.150$\\
\hline
8640& $0.9669 \times 0.7030$&9.132&$0.133$\\
    & $2.617 \times 1.996$&-79.15&0.266\\ 
\hline
\end{tabular}

\end{table}

\subsection{Radio Continuum Images}\label{sec:continuum}

\begin{figure*}
\includegraphics[scale=0.5,angle=270]{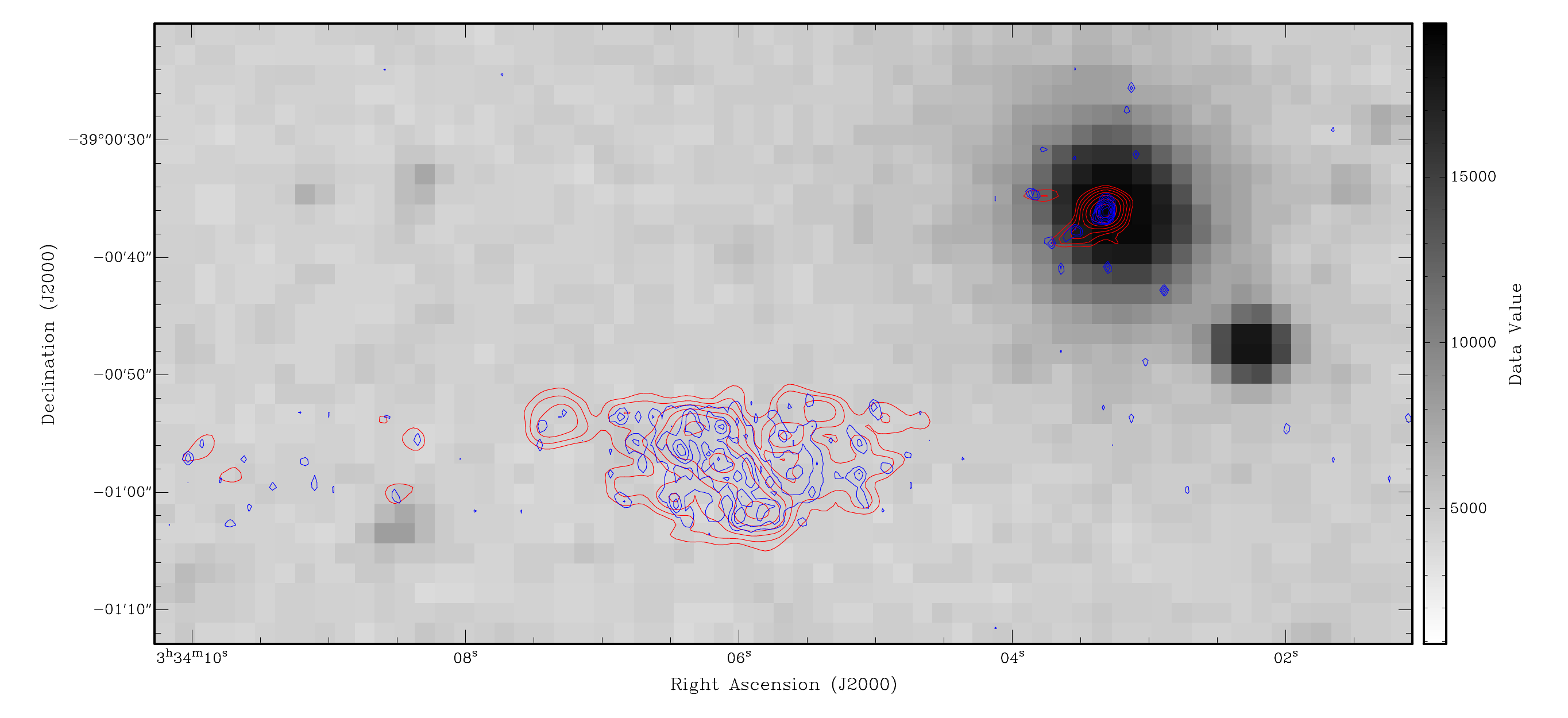}
\caption{Digitized Sky Survey blue image of the core of PKS J0334-3900 and the bottom knot overlaid with 8640 MHz Stokes-I contours. Red contours are from a tapered image (beam size $2.617^{\prime\prime} \times 1.996^{\prime\prime}$, RMS noise 0.266 mJy b$^{-1}$), while blue contours give the highest resolution imaging and correspond to a beam size is $0.9669^{\prime\prime} \times 0.7030^{\prime\prime}$ with RMS noise of 0.133 mJy b$^{-1}$. Contours start at 3 times the RMS noise and increase in multiples of $\sqrt{2}$. In each set of contours it is possible to observe a shape that looks like the bottom jet leaving the AGN. Note the presence of the close companion galaxy to the south west.}\label{fig:StartOfJet}
\end{figure*}

\begin{figure}
\includegraphics[width=7.5cm,angle=270]{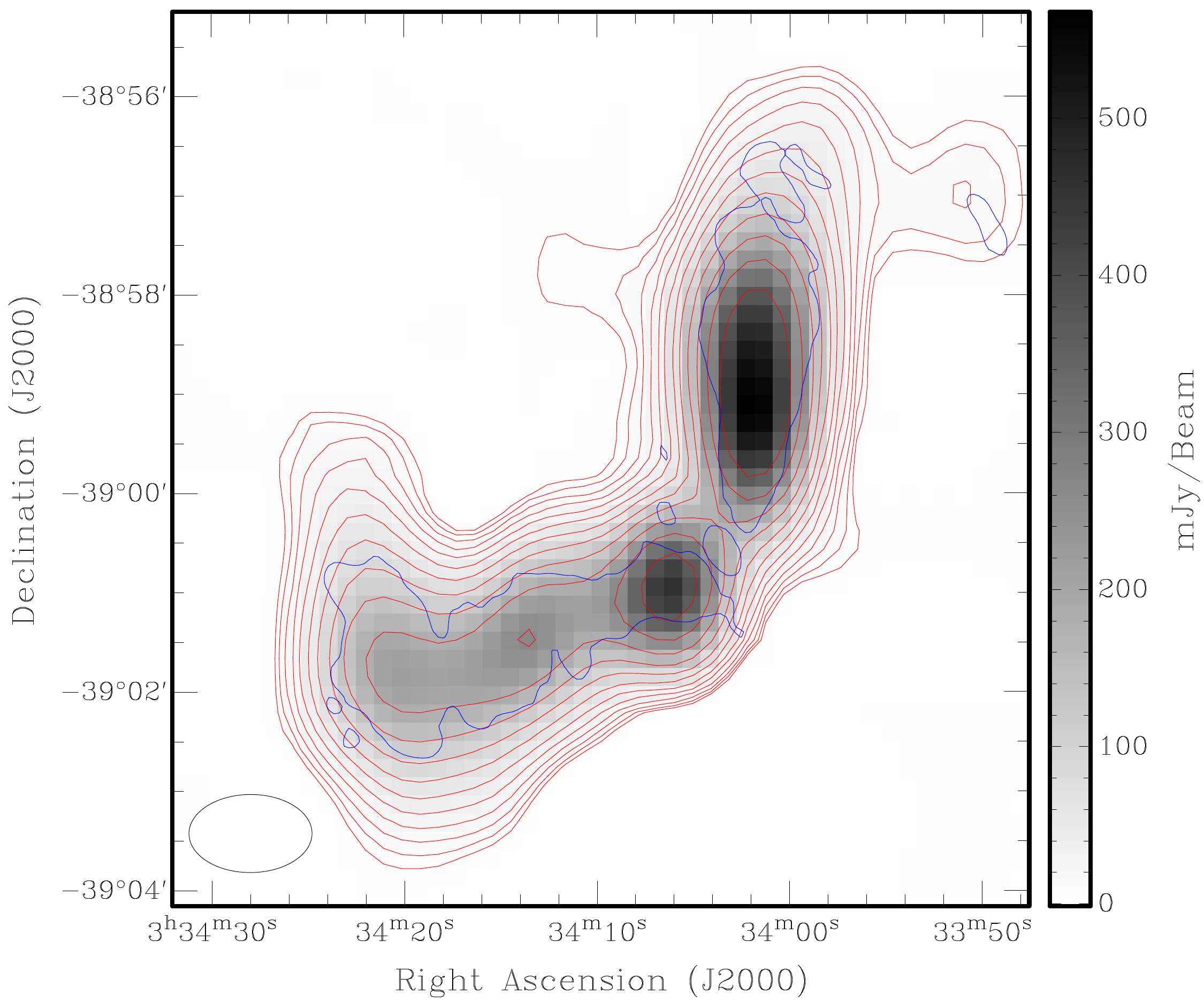}
\caption{MOST 843 MHz Stokes-I map with a beam size of 43" $\times$ 43". Red contours are the of the MOST 843 MHz Stokes I map, while the blue is the 3$\sigma$ contour from the tapered 1384 MHz ATCA map presented in this work (see Figure \ref{fig:cont1384}). At 843 MHz, there is more resolved low surface brightness emission at the ends of the jets, showing them to be longer than the 1384 MHz ATCA Stokes I image.}\label{fig:moststokesI}
\end{figure}

The Stokes I continuum images show PKS J0334-3900 to be bent at an obtuse angle, the top jet is directed north and the bottom jet is directed south-east, each jet has a bright knot close to the core and ends with a clear asymmetric hook morphology (see Figure \ref{fig:cont1384}). While there are similarities in shape between the jets, each jet varies differently in surface brightness. We find the jet's brightness variation is affected by its geometry, velocity, and orientation. For example, at 1.4 GHz (see Figure \ref{fig:cont1384}), the knot in the top jet is elongated in the jet's direction, but the knot in the bottom jet is brighter and is not as elongated. This can be explained if the bottom knot was directed towards the observer, it is expected to be compact in the plane of the sky and have a higher surface brightness. This explanation is also consistent with the 2.5 GHz observation (see Figure \ref{fig:cont2496}), the knot in the top jet has the same elongation, however, the knot in the bottom jet has newly resolved structure.

We now consider the surface brightness along the hooks and how this relates to their orientation. Figure \ref{fig:cont1384} shows the hooks of the top and bottom jets. Each hook has a local peak in intensity, the bottom jet has a peak at the start of its hook, and the top jet has a peak half way around its hook. After the relative peaks, the bottom jet intensity decreases steadily around the bottom hook, and the top jet decreases to the end of the hook. For each jet, the intensity starts decreasing at different locations around their hooks. This difference could be explained if there is a difference in orientation between the hooks along the line of sight. If the end of the bottom hook is directed away from the observer, this could explain its steady decrease in intensity. This would mean that the bottom jet starts off directed towards the observer and changes direction when it reaches the bend around the hook. This explanation would suggest that the orientation of the top jet does not change as drastically with respect to the plane of the sky.

Figure \ref{fig:StartOfJet} shows the Digitized Sky Survey (DSS) image overlaid with Stokes I 8.6 GHz contours. At 8.6 GHz, the start of the bottom jet is detected at the radio core. The lack of a counter jet and the angle the bottom jet leaves the core suggests that the bottom jet is in the foreground \citep{parma94}. This detection is not visible at any of the other frequencies due to lack of resolution. Next to the core elliptical galaxy is a nearby companion galaxy (see Figure \ref{fig:StartOfJet}), with an apparent Johnson blue magnitude of b$_J=17.41 \pm 0.08$  \citep{hambly01b,hambly01a}. This companion has a projected distance of 20.54 kpc from PKS J0334-3900. 

Radio observations of PKS J0334-3900 by MOST (843 MHz) (see Figure \ref{fig:moststokesI}) \citep{bock99} and the NVSS (1.4 GHz) show a similar morphology, while being at lower resolution. Unsurprisingly, the MOST observation has more diffuse low surface-brightness emission which cannot be seen in the 1384 MHz ATCA observation (see Figure \ref{fig:moststokesI}), this is likely due to the spectral steepening of the source. Since the MOST observation had more low surface brightness emission, we used it to measure the projected linear size of each jet (see Table \ref{tab:SpixInfo}).

The physical attributes of PKS J0334-3900 were measured and can be found in Table \ref{tab:SpixInfo}, this includes values of spectral index, flux, and total power. The uncertainties in the flux were calculated using ${\Delta}S=\sqrt{a^2 +(bS)^2}$, where $a$ is the noise of the image, $b$ is the residual calibration error (assumed to be 0.01 \citep{venturi00}), and $S$ is the flux.

 At 1384 MHz, PKS J0334-3900 has a total power of $1.46 \times 10^{25}$ W Hz$^{-1}$, this places it at the transition phase between being an FR I and II source \citep{ledlow96}, which is typical for HT galaxies.

\begin{table*}
\caption{Details of measured total length along the jets, flux and spectral index for the core, top jet, bottom jet and over the whole HT galaxy. The length was measured using the MOST 843 MHz observation. Linear least squares fits have been applied to determine spectral indices. Literature values used for the entire galaxy are shown in Table \ref{tab:AllSpixValues} and the core values are drawn from \citep{slee94}. }\label{tab:SpixInfo}
\begin{tabular}{l l l l l l l l}
\hline
Jet & Linear Size &Frequencies&Flux & Polarized Flux & $\alpha$ This work& $\alpha$ Inc. literature& P\\
&(kpc) & MHz & (mJy)& (mJy) & & & W$Hz^{-1}$\\ 
\hline
Total & 564 & 1384 & $1913 \pm 22$ & $280 \pm 3$ &$-0.79 \pm 0.01$&$-0.72 \pm 0.01$& $1.46\times 10^{25}$\\
& & 2496 & $1368 \pm 16$& $190 \pm 1$& & &$1.05\times 10^{25}$\\
& & 4800 & $937 \pm 12$&$28 \pm 0.2$& & &$7.17\times 10^{24}$\\
& & 8640 & $445 \pm 5$& & & &$3.40\times 10^{24}$\\
\hline
Top & 263 & 1384  & $785 \pm 9 $&$90 \pm 0.9$&$-0.71 \pm 0.01$\\
& & 2496 &$697 \pm 8$& $100 \pm 1$\\
& & 4800 &$438 \pm 5$& $10 \pm 0.1$\\
& & 8640 &$222 \pm 2$\\
\hline
Bottom & 301 & 1384 & $1120 \pm 13$  &$190 \pm 2$&$-0.88 \pm 0.01$\\
& & 2496 & $841 \pm 9$ &$ 90 \pm 0.1$\\
& & 4800 & $488 \pm 7$& $14 \pm 0.1$\\
& & 8640 & $210 \pm 2$\\
\hline
Core &  & 1384 &$8 \pm 0.8$ & &$-0.15 \pm 0.02$&$-0.15 \pm 0.02\*$\\
& & 2496 & $11 \pm 0.3$ &  & & \\
& & 4800 &$12 \pm 0.2$&  & & \\
& & 8640 &$13 \pm 0.2$& & & \\
\hline
\end{tabular}
\end{table*}

\subsection{X-ray Analysis}
\label{chandra}
\begin{figure}
\includegraphics[width=80mm,angle=-90]{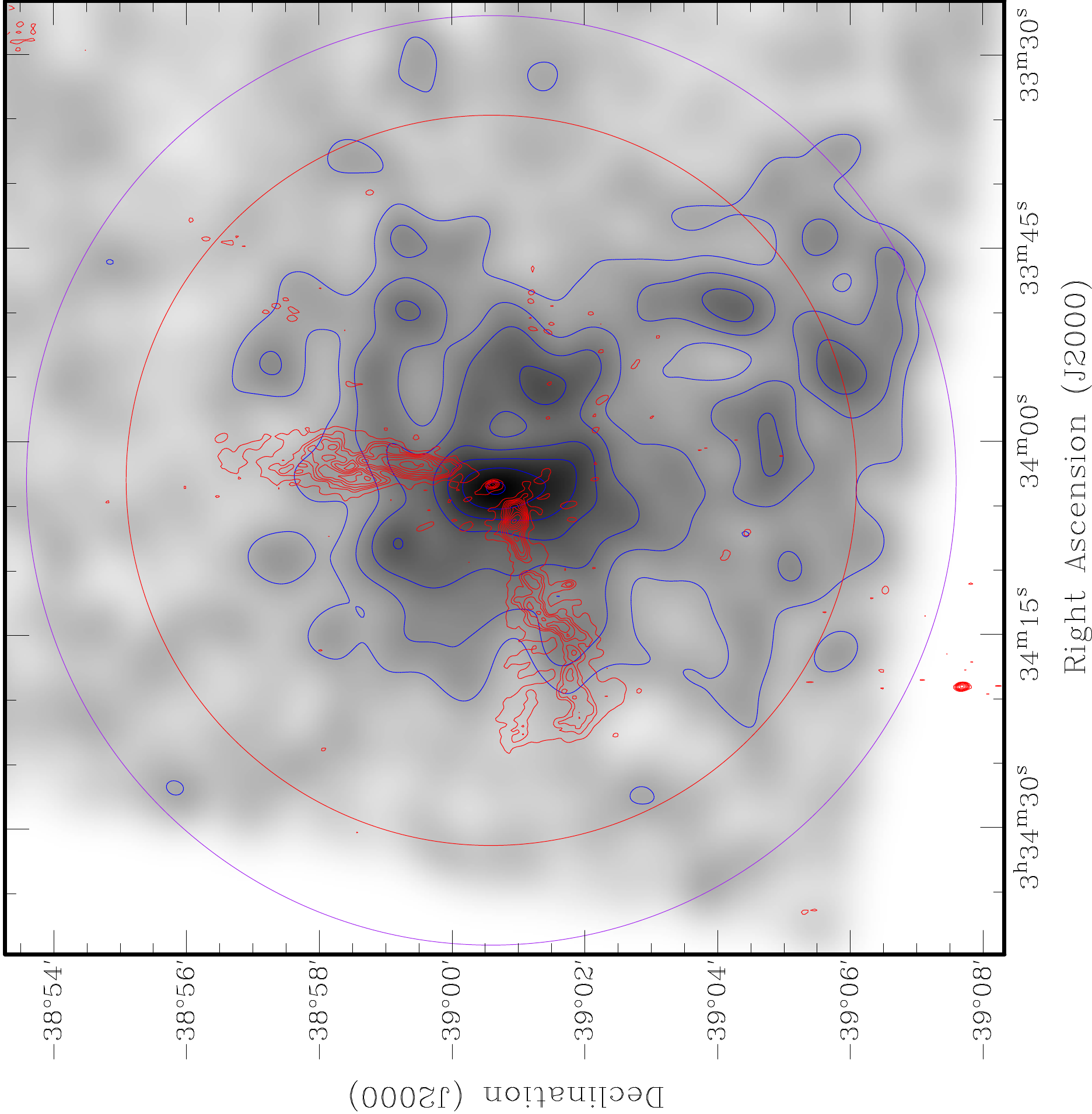}
\caption{Chandra observation of A3135 which has been adaptively smoothed. The image has been corrected for background counts and exposure, and point sources removed. The blue contours are at the values 0.018, 0.024, 0.030, 0.036, 0.042, 0.048 \& 0.054 and the red shows the 1.4 GHz radio image of PKS J0334-3900. The red circle has a radius of 5.5$^{\prime}$, within which we have a reasonable kT temperature. The calculated estimate of $n_e$ was determined inside the purple circle of radius 7$^{\prime}$.}\label{fig:chandra}
\end{figure}

The X-ray observation of Abell 3135 was performed with the Advanced
CCD Imaging Spectrometer (ACIS) on 2009 May 7
(obsID: 9393). Standard Chandra data analysis was performed
which includes the corrections for the slow gain change,
charge transfer inefficiency and the ACIS optical blocking filter 
contamination. The observation {was filtered for background flares and 
none were found}. The total exposure is 15.3 ks.
We used CIAO4.4.1 for the data analysis. The calibration files used
correspond to Chandra calibration database (CALDB) 4.5.0
from the Chandra X-ray Center. 
The solar photospheric abundance table by \cite{anders89} is used in the spectral 
fits. Uncertainties quoted in the X-ray analysis are 1$\sigma$. 
The 0.7 - 2 keV Chandra image of A3135, with background subtracted and exposure corrected, is shown in Figure \ref{fig:chandra}.
X-ray point sources have also been removed and filled with the average of the surroundings. The shown image was 
adaptively smoothed with the Science Analysis System (SAS) tool ASMOOTH. 
Default parameters of ASMOOTH \footnote[1]{http://xmm.esac.esa.int/sas/current/doc/asmooth/index.html} were used, which includes a
maximum width of the gaussian kernel of 19.68$''$ (or 40 ACIS pixels)
and a desired signal-to-noise ratio of 10.

\begin{figure}
\includegraphics[width=85mm]{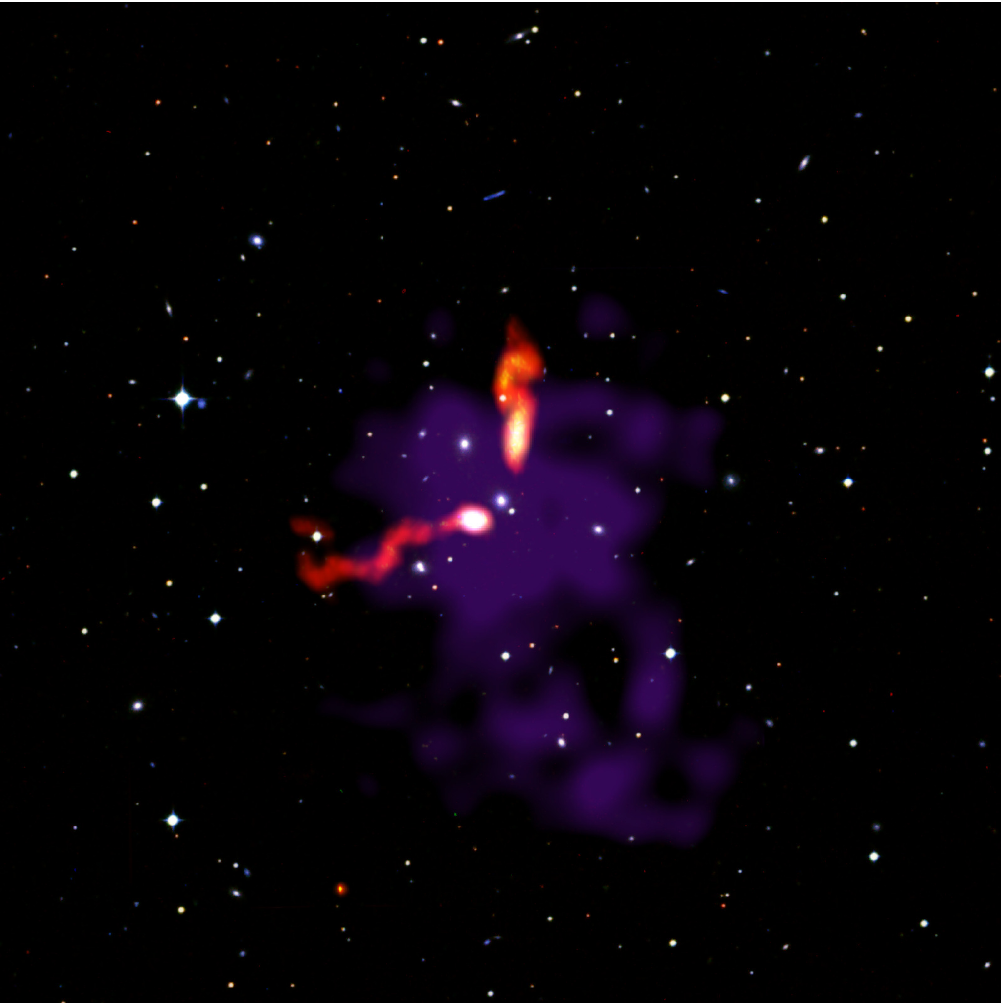}
\caption{Three colour (blue, red, and infrared) image of A3135 overlaid with Chandra observations (purple) and 1.4 GHz ATCA observations. The image colour scales have been adjusted in a similar manner to images made for the Hubble Heritage Project.}\label{fig:pretty}
\end{figure}

For the spectral analysis, we use a Galactic absorption column
density of $1.6 \times10^{20} \textrm{cm}^{-2}$ \citep{kalberla05}.
The global spectrum is extracted within a radius of 5.5$'$ and the
point sources are excluded.  We followed the method
described in \citet{sun09b} for background subtraction,
which involves the use of the ACIS stowed background and the modeling of
the cosmic X-ray background. After the removal of the particle background and
the cosmic X-ray background, there are 1349 counts (0.5 - 7.0 keV) from the
cluster in the global spectrum. The spectral fitting was performed 
with XSPEC 12.7.1 and the spectral model is APEC.
The measured global temperature and
abundance are 2.41 $^{+0.51}_{-0.38}$ keV and 0.08 $^{+0.15}_{-0.08}$ solar
respectively for AtomDB 1.3.1. The results with AtomDB 2.0.1 are
very similar. C-statistic does not provide a goodness-of-fit measure but the XSPEC
version of the C-statistic is defined to approach $\chi^{2}$ in the case of
large number of counts. The C-statistic of the fit is quoted for reference,
62.0 for 53 degrees of freedom. 

The radial surface brightness profile of the cluster, 
centered on the central galaxy, is also measured in the 0.7 - 2.0 keV band. The 
profile is analyzed with the CIAO/Sherpa tool and it is well fitted by a $\beta$ model.
Combined with the results of the global spectral fitting, the radial
electron density profile of the cluster is derived:

\begin{align*}
n_e &= n_{e,0} [1 + (r/r_s)^2]^{-1.5\beta},\\
n_{e,0} &= (1.06 \pm 0.11) \times 10^{-3} \textrm{cm}^{-3} ,\\ 
\beta &= 0.504 \pm 0.016, \\
r_s &= 133 \pm 27 \textrm{kpc} .
\end{align*}

Figure \ref{fig:chandra} shows the cluster X-ray emission is skewed to the southwest.
There are several large elliptical galaxies around the cluster center,
including one (AM 0332-391 NED02, ${\rm RA_{J2000}}$ = 03:34:06.7,  ${\rm DEC_{J2000}}$ = -38:59:33)
that is $\sim$0.13 mag brighter than PKS J0334-3900 at the K$_s$ band.
It is likely that the system is in a late merging stage and the
merger happened along the SW-NE direction on the plane of sky,
which is consistent with the elongated optical distribution shown
in Figure \ref{fig:RedShiftDist}. 

Previous X-ray analysis has shown PKS J0334-3900 is associated with a small 
X-ray cool core (or a corona), (see \citet{sun09}, Table 2). There are no 
significant cavities detected in the X-ray emission. However, we can
use the volume of the radio lobes to estimate the cavity power, e.g. as 
in \cite{birzan08}. If we assume a prolate spheroid for each lobe with 
140 kpc $\times$ 40 kpc $\times$ 40 kpc for both the upper lobe and
the lower lobe, using the average values for $n_e (6 \times 10^{-4} \textrm{cm}^{-3}$) and kT (2.4 keV),
we find 4PV $\sim 10^{60}$ erg.
Assuming the sound speed for the bubble rising speed \citep{mcnamara07},
the cavity power is then $\sim$ 1.6 $\times 10^{44}$ erg s$^{-1}$,
which is reasonable for PKS J0334-3900's radio luminosity.
This value can also be taken as a lower limit for the jet power.

We present the Chandra data in combination with both radio and a three colour (blue, red, and infrared) image in Figure \ref{fig:pretty}. This image is presented to
give a larger scale view of all emission in this system and provide an aesthetically pleasing example of the complexities of cluster emission.

\subsection{Spectral Indices}
 Generally, the synchrotron spectrum of radio galaxies follows the power law $S \propto \nu^\alpha$, where $S$ is flux, $\nu$ is frequency, and $\alpha$ is the spectral index. The spectral index is useful as it determines the synchrotron spectrum for the source, and when used with the critical frequency, can be used to estimate the age of the synchrotron emission.

\begin{figure}
\centering
\includegraphics[width=9cm]{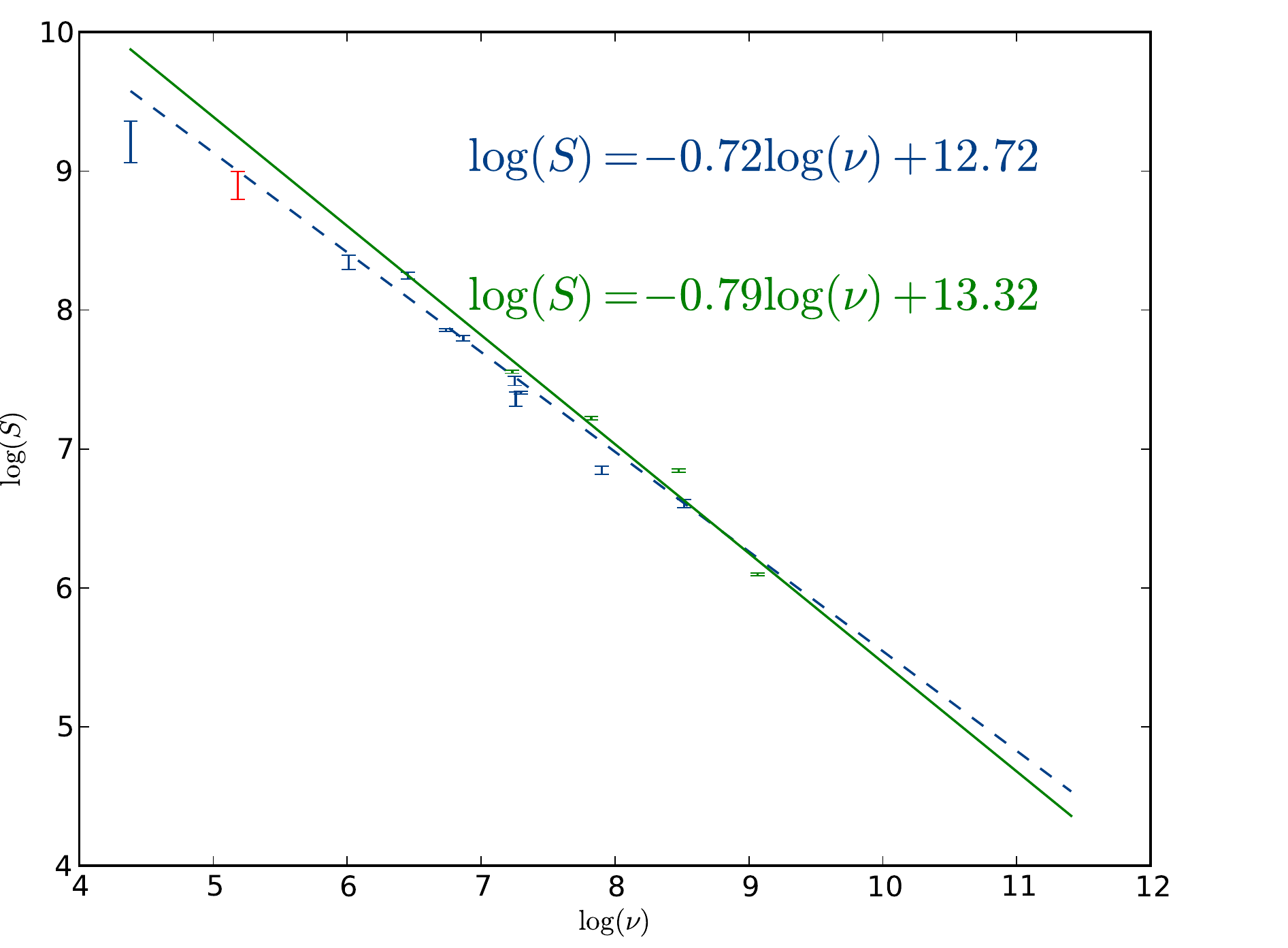}
\caption{Total spectral index plot of PKS J0334-3900 using only the ATCA data from this work (green), and with all available values from literature (blue). The equation of the solid trend-line for the ATCA only data is $\log{S}=-0.79\log{\nu} + 13.32$, while the ATCA and literature values produce the dashed trend-line, with the equation $\log{S}=-0.72\log{\nu} + 12.72$. The ATCA only line corresponds to data in Table \ref{tab:SpixInfo}, whereas the literature values along with the ATCA values from this work can be found in Table \ref{tab:AllSpixValues}. Note that the red data point represents an interpolated value from literature and was not used in the line fitting.}
\label{fig:ThisWorkAverageSpix}
\end{figure}

Flux values for PKS J0334-3900 were measured from this work and collected from the literature, in order to determine the spectral index. These values were plotted as $\log S$ vs $\log\nu$, where the gradient is the spectral index. The total spectral index of the source from the ATCA data (see Figure \ref{fig:ThisWorkAverageSpix}) has a value of $\alpha$ = -0.79 $\pm$ 0.01. 
The literature has a large number of flux densities recorded for PKS J0334-3900 across an extremely wide range from 80 MHz to 8.6 GHz (see Table \ref{tab:AllSpixValues}). These literature values were combined with the measured values in this work (see Figure \ref{fig:ThisWorkAverageSpix}), and the spectral index of the source was calculated to be $-0.72\pm 0.01$. 

Not all of the literature values were used in this calculation, due to three factors. First, the flux density for 160 MHz has been left out of this calculation, since the low resolution suggests nearby sources are blended with PKS J0334-3900, making it unreliable as a flux estimate. Second, where there are multiple flux densities at the same frequency, we selected the most reliable value for the calculation (see Table \ref{tab:AllSpixValues}). Finally, the 178 MHz data point was not included in the calculation (seen as red in Figure \ref{fig:ThisWorkAverageSpix}), because it is an interpolated value. 

Figure \ref{fig:ThisWorkAverageSpix} shows that the point at 80 MHz appears to show some deviation from the power law suggesting the critical frequency to be near the interpolated point at 178 MHz. If we assume 178 MHz is the critical frequency, we may use the method deployed in \cite{lal04} to calculate an estimated upper limit age of the synchrotron emission using equation (1) (the details behind this calculation are given in \cite{miley80}).  

\begin{align}
& &t = 1060\frac{B^{1/2}}{B^2+B^2_{IC}}(1+z)^{-1/2}{\nu_{br}}^{-1/2}
\end{align}

where t is the upper limit to the age of the jets in Myr. B and $B_{IC}$ are the total and inverse Compton derived magnetic fields respectively, both measured in $\mu$G, and $\nu_{br}$ is the break frequency of the injected electron population in the jets.  

This method assumes the magnetic field magnitude for the inverse Compton component is given by
\begin{align}
& &B=\frac{B_{IC}}{\sqrt{3}} 
\end{align}
and is a function of redshift and can be estimated as $B_{ic}=4(1+z)^2$ for the case of maximum radiative lifetime of the particles. These equations yield a magnetic field value of $B=2.6 {\mu}$G and an upper limit age of 146 Myr for the jet emission.

\begin{table}
\caption{All frequencies and net flux densities of PKS J0334-3900 currently known, along with ones measured in this paper. $\star$ this is not a measured value, but rather is an interpolated value between literature low frequency measurements. We list the interpolated value here for completeness, but do not use it in calculations of the total source spectral index. We have assumed a 10 per cent error for this interpolated source. $\circ$ claimed in the literature to be from \citet{schilizzi75}, the correct value is not found in that paper but rather in \citet{wall79}. $\ast$ measured from the Sydney University Molonglo Sky Survey image. $^{\dagger}$ measured from the NVSS image. There is a corresponding graph, Fig. \ref{fig:ThisWorkAverageSpix}, that includes all the frequencies denoted with a $^1$, but does not include the flux density at 160 MHz as this was believed to be unreliable.}\label{tab:AllSpixValues}
\begin{tabular}{l l l}
\hline
Frequency & Flux Density & Reference\\
MHz& mJy \\
\hline
$80^1$&	$10000 \pm 1500$& \cite{mills60} \\
$160$&	$3700 \pm 555$& \cite{ekers89} \\
178&        7300& \cite{burgess06}$^{\star}$ \\
$408^1$&	$4200 \pm 210$& \cite{wall79}$^{\circ}$ \\
$635^1$&	$3820\pm 96$& \cite{wills75} \\
843  &       2410& \cite{jones92}\\
$843^1$&	$2578\pm 25$& This work$^{\ast}$\\
$960^1$&	$2430\pm 48$& \cite{wills75}\\
$1384^1$&$1913\pm 23$& This work\\
1410 &      1700& \cite{bolton65}\\
$1410^1$ &      $1790 \pm 60$& \cite{wills75}\\
$1420^1$&	$1570\pm 78$& \cite{ekers89}\\
$1477^1$&	$1642 \pm 16$&  This work$^{\dagger}$\\
$2496^1$&	$1368\pm 16$& This work \\
2700&       900& \cite{bolton73}\\
$2700^1$&	$940\pm 28$& \cite{wall73}\\
$4800^1$&	$937\pm 12$& This work\\
$5000^1$&	$740\pm 22$& \cite{wall79} \\
$8640^1$&	$445\pm 5$& This work\\
\hline
\end{tabular}
\end{table}

\subsection{Polarisation}
\begin{figure*}
\includegraphics[width=190mm]{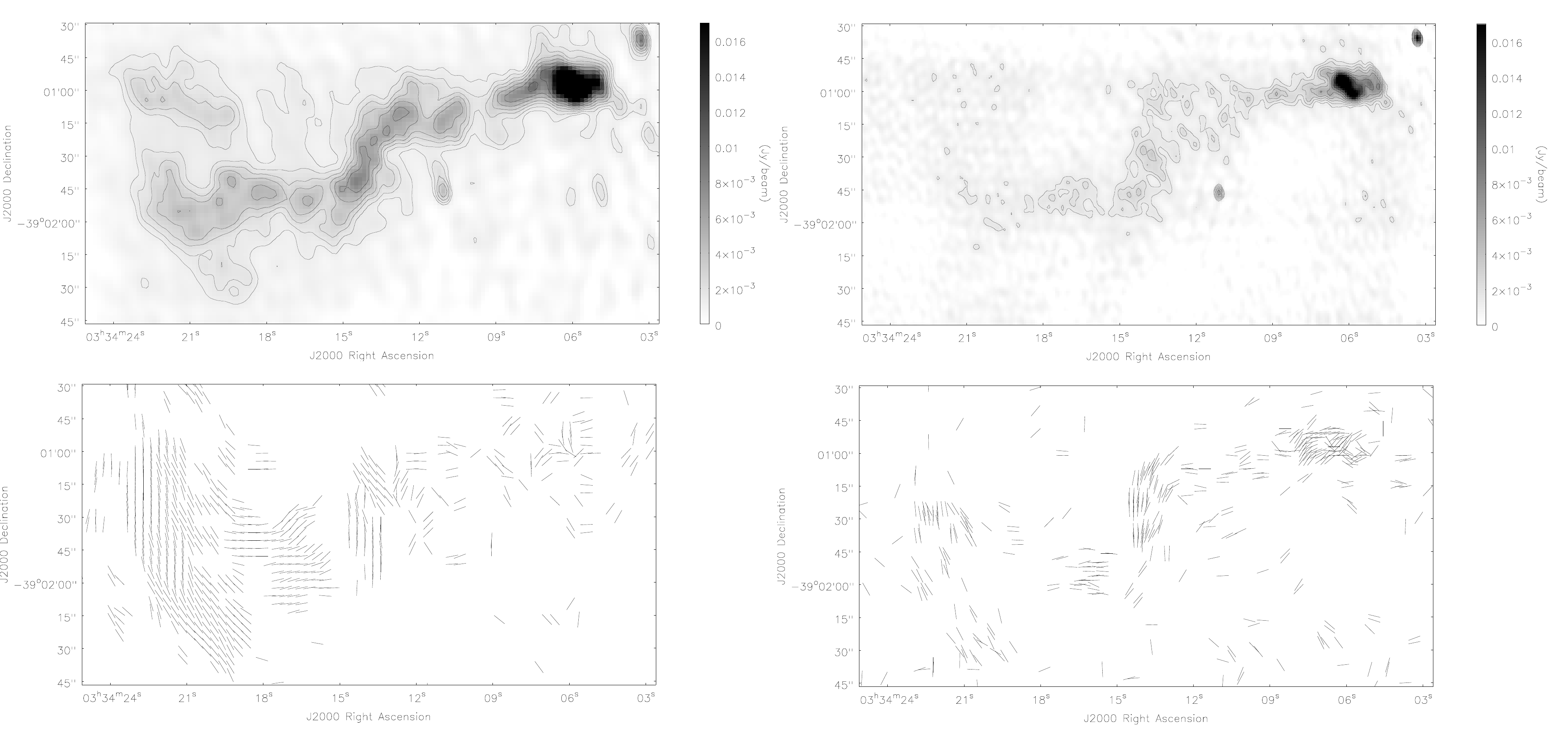}
\caption{The direction of the magnetic component of the polarised emission is compared to the morphology of the bottom jet. The lefthand images are at 1384 MHz, and the righthand images are at 2496 MHz. The contours on the intensity images start at $3\sigma$ then increase as multiples of $\sqrt{2}$. The image details are given in Table \ref{tab:pol}. Note the downward `finger' of emission in the 1384 MHz image at ${\rm RA_{J2000}}$ 03:34:11.06 is due to the AGN associated with cluster member AM 0332-391 NED04, which is clearly seen as a discrete source in the 2496 MHz image on the righthand side.}\label{fig:polarisationbottom}
\end{figure*}

\begin{figure*}
\includegraphics[width=190mm]{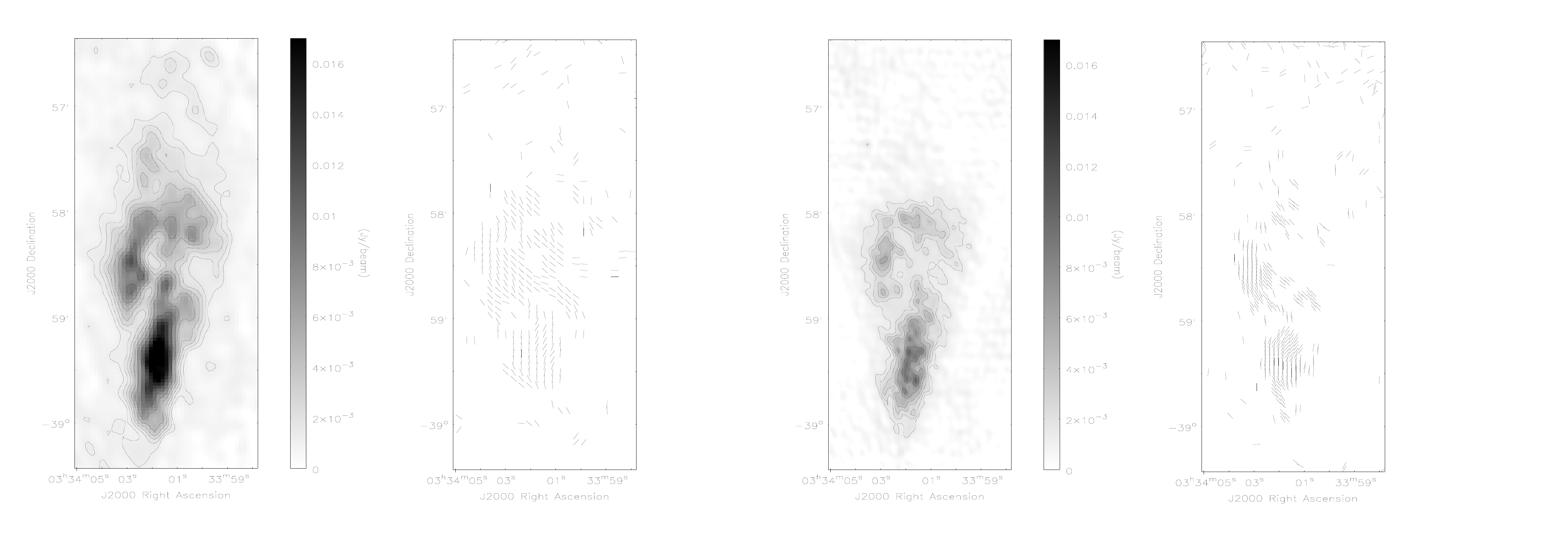}
\caption{The direction of the magnetic component of the polarised emission is compared to the morphology of the top jet. The first two images on the left are at 1384 MHz, and the last two on the right are at 2496 MHz. The contours on the intensity images start at $3\sigma$ then increase as multiples of $\sqrt{2}$. The image details are given in Table \ref{tab:pol}. }\label{fig:polarisationtop}
\end{figure*}

\begin{figure}
\includegraphics[width=6cm, angle =-90]{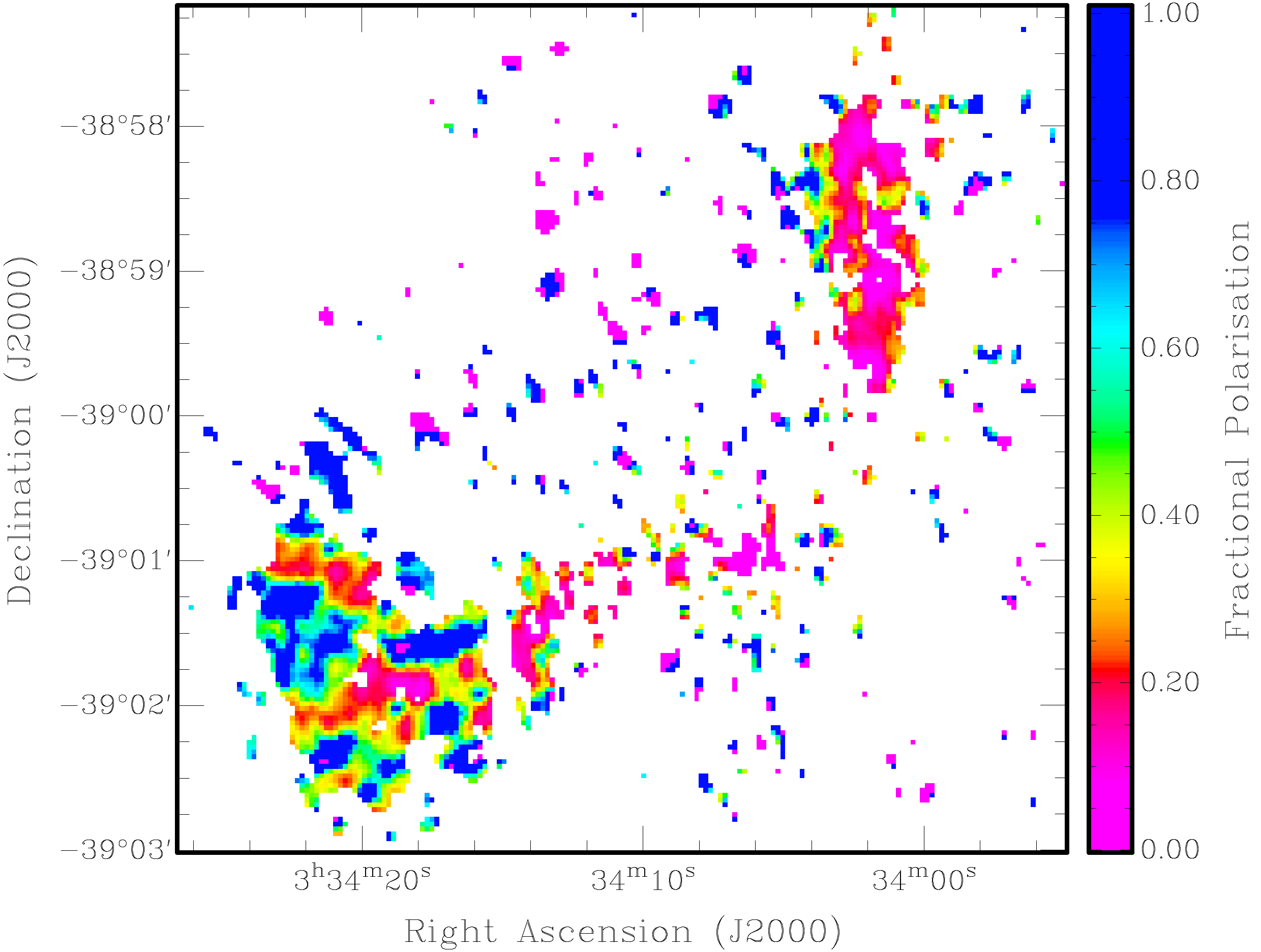}
\caption{Fractional polarisation image at 1384 MHz showing a much higher percentage in polarisation in the bottom jet as compared to the top. The image details are given in Table \ref{tab:pol}.}\label{fig:fractionalpolraisation}
\end{figure}

Stokes Q and U images of PKS J0334-3900 were made from the 1384 and 2496 MHz observations, and used to generate maps of the linear polarisation angle (PA) and the fractional polarisation (see Figures \ref{fig:polarisationbottom}, \ref{fig:polarisationtop} \& \ref{fig:fractionalpolraisation} and Table \ref{tab:pol}). These maps are useful for understanding the jet morphology. The PA maps have their angles rotated by $90^\circ$ to show the direction of magnetic field along the jets. 

Figures \ref{fig:polarisationbottom} \& \ref{fig:polarisationtop} show the magnetic field (B) vectors along the bottom and top jets respectively. The PA values are masked if their error was greater than $10^\circ$, or if the polarised intensity was below 3 sigma. At each frequency the vectors point along the jets. 

At the start of the bottom jet (see Figure \ref{fig:polarisationbottom}), the 2.5 GHz observations show the magnetic field leaving the knot, and curving steeply around the start of the hook. This steep curvature of the magnetic field corresponds to where there is a peak in intensity. This may suggest that the jet is directed towards the observer just before curving into the hook. 
Just after the hook starts, there is a break in the polarisation signal at ${\rm RA_{J2000}}$ = 03:34:14. The break may be a sign of the jet changing direction. At 1.4 GHz, the vectors are perpendicular to the break at the start of the hook, and then follow the curve of the hook thereafter. However, at 2.5 GHz, the vectors appear to be curving into the hook just before the break. If the break in polarisation is where the hook changes direction along the line of sight, it is consistent with the surface brightness decreasing around the hook, as explained in Section \ref{sec:continuum}. 
At the end of the hook, there is a large, ordered field structure and high fractional polarisation (see Figure \ref{fig:fractionalpolraisation}). Additionally, the bottom jet has significantly higher fractional polarisation when compared to the top jet.  The 2.5 GHz observation suggests that the end of the bottom jet could be curving down, however the 1.4 GHz image shows that the situation is more complicated, suggesting upward motion. However, it is well known from the Laing-Garrington effect \citep{laing88,garrington88} that the foreground jet in a radio galaxy will exhibit higher fractional polarisation, suggesting the bottom jet and in particular the end of the jet is directed towards the observers, and thus is likely to be moving up and towards the line of sight.

The top jet has a curved flow out of the knot and bends around the hook of the top jet (see Figure \ref{fig:polarisationtop}). The intensity image does not show the jet curving out of the knot, however, it is consistent with the observed hook shape. At the end of the hook shape, the 1.4 GHz vectors suggest that the jet flows down at a 45 degree angle. The lack of PA vectors at the end of the jet make it difficult to interpret the flow.

\begin{table*}
\caption{Details of continuum Stokes I, Q and U images shown in Figures \ref{fig:polarisationtop} \& \ref{fig:polarisationbottom}.}
\begin{tabular}{l l l l l l}
\hline
Frequency &Synthesized Beam Size & PA & $\sigma$(I)& $\sigma$(Q)&$\sigma$(U) \\ 
(MHz)&(arcseconds)&$\circ$ &(mJybeam$^{-1}$)&(mJybeam$^{-1}$)&(mJybeam$^{-1}$) \\ 
\hline
1384\\
Polarised Image& $8.749 \times 5.044$  &9.738& &$0.174$ &$0.122$ \\
Intensity Image& $15 \times 15$ &0&$1$&& \\
\hline
2496\\
Polarised Image& $4.977 \times 2.871$ &7.897&&$0.197$&$0.192$\\
Intensity Image& $15 \times 15$ &0& $0.833$ &&\\
\hline
\end{tabular}
\label{tab:pol}
\end{table*}

\subsection{Rotation Measures}\label{sec:RM}
Faraday rotation is the phenomenon of linearly polarised light travelling along a magnetic field causing a rotation in polarisation angle ($\chi$). This rotation depends on the electron density ($n_{e}$), magnetic field magnitude along the line of sight ($B_{\parallel}$), distance passed through the magnetic field, and the wavelength of the linearly polarised light ($\lambda$). 
Rotation measure, $\rm{RM} = \frac{d\chi}{d\lambda^2}$, follows the relationship $\rm{RM} \propto \int n_{e}\vec{B_{\parallel}}\cdot\ d\vec{r}$. The value of $\rm{RM}$ can be obtained by measuring $\chi$ for different values of $\lambda$. As tailed radio galaxies present extended linearly polarised structures, they have been used in studies of cluster magnetic fields using Faraday rotation \citep{clarke01,eilek02,vogt03,johnston-hollitt03,guidetti08}.

 Using two different methods, the rotation measure across PKS J0334-3900 was calculated 
 using the 1384 MHz Stokes Q and U observations. The first method was rotation measure synthesis \citep{brentjens05}, and the second the traditional measurement of the gradient of the PA with frequency \citep{burn66}. The RM values from these methods were then compared. Finally, a RM synthesis map of PKS J0334-3900 was made, and using the vector averaged polarisation an RM for each jet was calculated. 

The RM values were measured within a bandwidth of 96 MHz, using 8 MHz intervals. Each interval had a continuum Stokes Q and U map which were combined into cubes. Only 12 intervals were used, because the interval centred at 1408 MHz was removed due to cross-channel harmonics. The Stokes Q and U maps (planes in the cube) have values of RMS noise between 330 \& 380 nJy, and the Stokes Q and U RMS values were added in quadrature to obtain the RMS noise in linear polarisaton. 

Values of RM\textsubscript{Linear} were calculated using the method of a least squares fit of $\chi$ vs. $\lambda^2$. $\lambda^2$ was determined by the mean frequency of each interval. Values of RM\textsubscript{Linear}  were masked when 5 or more of the points in $\lambda^2$-space had linear polarisations less than $3\sigma$. $\chi$ values were corrected by $n\pi$  where there was an obvious ambiguity. As the intervals are closely spaced, there were no instances where more than a single $\pi$ correction needed to be applied. 

\begin{figure}
\centering
\includegraphics[width=7.5cm]{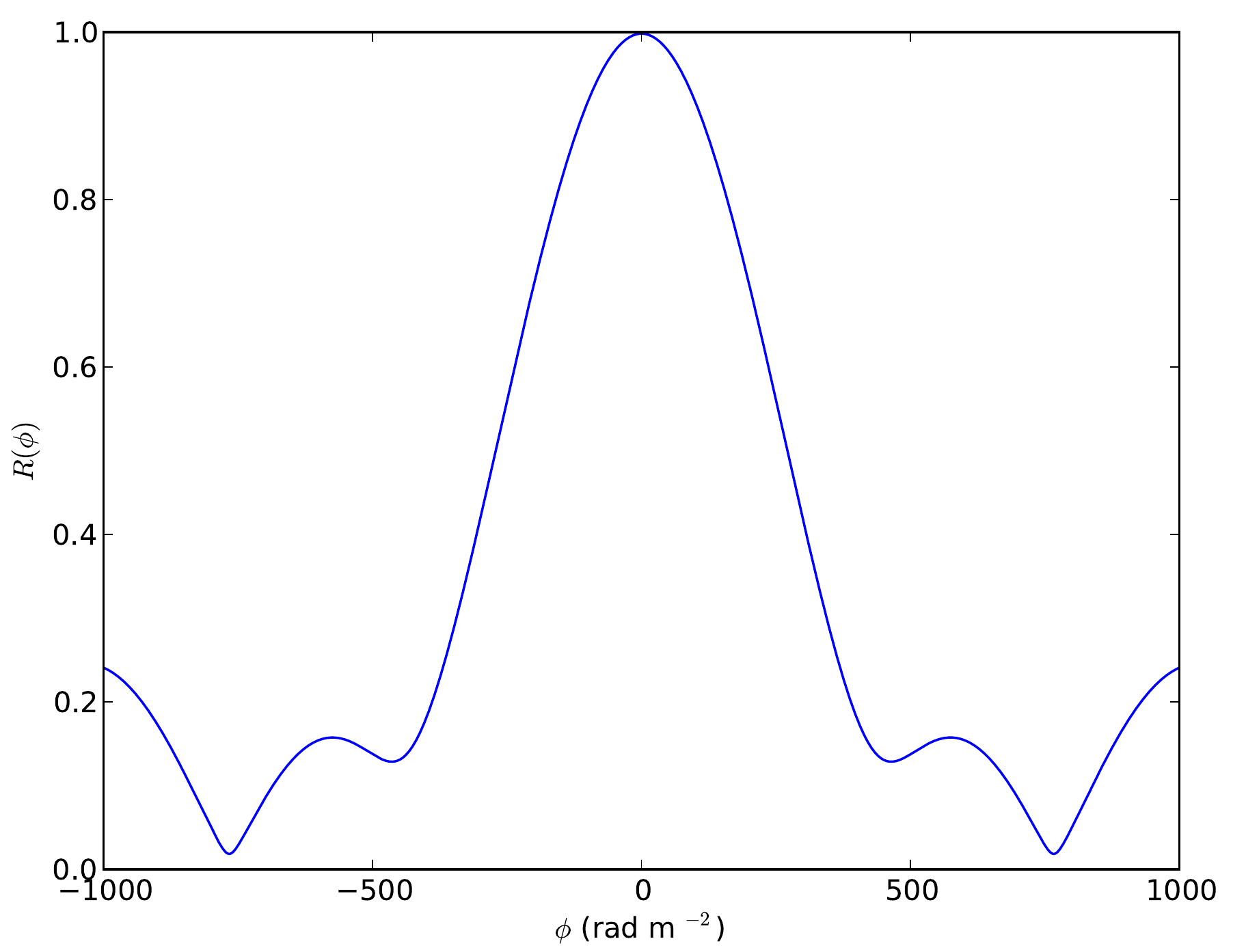}
\caption{The Rotation measure transfer function over the range [-1000,1000] rad m$^{-2}$. This function has a full width at half maximum of 542.4 rad m$^{-2}$.}
\label{fig:RMTF}
\end{figure}

\begin{figure*}
\includegraphics[width=9.8cm,angle=270]{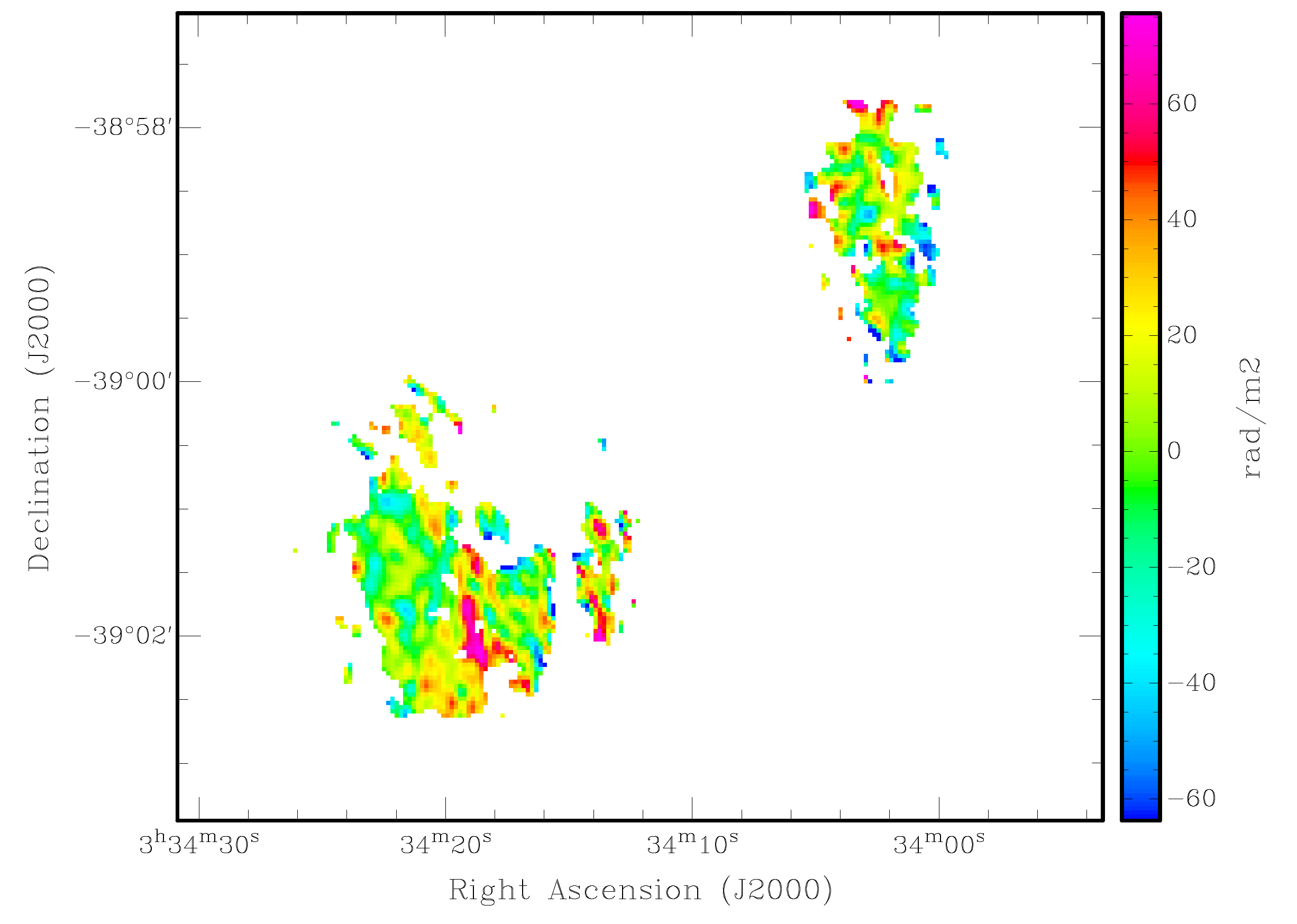}
\caption{Masked rotation measure image of PKS J0334-3900 made using rotation measure synthesis. The mask corresponds to the same mask as on the polarization angle images in Figure \ref{fig:polarisationbottom} and Figure \ref{fig:polarisationtop}  . The beam size of this rotation measure image is $7.1^{\prime\prime} \times 7.1^{\prime\prime}$. }\label{fig:rotationmeasure}
\end{figure*}

Values of RM\textsubscript{Synthesis} were measured using the RM synthesis technique, from the same 8 MHz intervals. These intervals were used to calculate the rotation measure transfer function (RMTF), shown in Figure \ref{fig:RMTF}. This RMTF has a full width at half maximum (FWHM) of 542.4 rad m$^{-2}$. The FWHM of the RMTF is large, so no attempt was made to deconvolve between components in RM\textsubscript{Synthesis}, because it is unlikely that the components would be resolved. Since multiple components could not be resolved, any possible magnetic fields in the foreground of the cluster cannot be distinguished. The RM\textsubscript{Synthesis} values were measured by finding the location of the peak in the Faraday dispersion function. This was done by fitting a 2nd order polynomial to the peak over an interval of $\pm100$ rad m$^{-2}$. These RM values were then used to generate an RM\textsubscript{Synthesis} map of PKS J0334-3900, shown in Figure \ref{fig:rotationmeasure}.

\begin{figure}
\centering
\includegraphics[width=8cm]{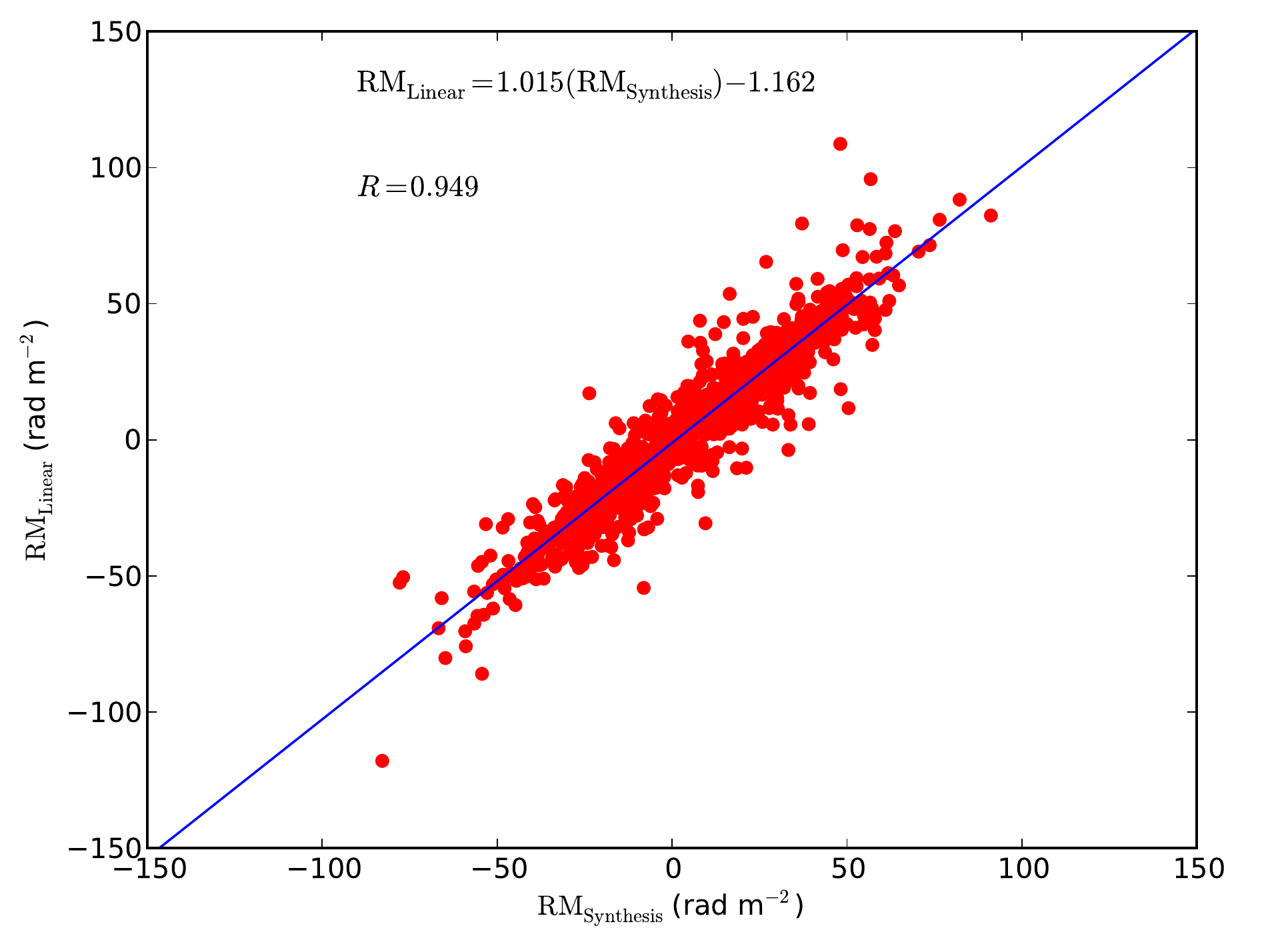}
\caption{Comparison of measurements of rotation measure generated via rotation measure synthesis and linear fits per pixel for PKS J0334-3900. The good agreement between the two methods is suggestive of a single component in RM space, and provides confidence that the methods were applied correctly.}
\label{fig:DiffScatterplot.eps}
\end{figure}

To compare the two RM methods for each pixel value a scatter plot of RM\textsubscript{Linear} vs. RM\textsubscript{Synthesis} was made  (see Figure \ref{fig:DiffScatterplot.eps}). The RM\textsubscript{Synthesis} values are expected to yield superior results to a straight linear fit in $\lambda^2$-space with less spurious results \citep{macquart12}. This scatter plot conforms closely to the expected RM\textsubscript{Linear}$=$ RM\textsubscript{Synthesis} relationship for a single source of RM along the observer's line of sight. This shows that the two methods are consistent on a pixel by pixel basis, as is expected for 97 per cent of sources with signal-to-noise ratio in excess of 3$\sigma$ in Stokes Q and U \citep{macquart12}. Additionally, the close agreement between the two methods demonstrates that the $n\pi$ ambiguity correction applied to the RM\textsubscript{Linear} values was reasonable.

The pixel by pixel RMs show good agreement with a $\lambda^2$ fit in the linear plots (not shown) and have only a single, albeit wide, peak in the RM synthesis results. Additionally, there is little depolarisation across the jets. Both the good $\lambda^2$ fit and the lack of depolarisation suggest that the RM measured is not from the jets of the tailed radio galaxy itself, but rather the tailed radio galaxy is providing polarised light which passes through a foreground Faraday screen. The reasoning for this is as follows: 1) if the RM was due to the jets, there would be different Faraday screens due to electron density and magnetic field variation through the interior of the jet, meaning there would not be a linear $\chi$ vs. $\lambda^2$ fit. 2) if the RM was from the jets, we expect depolarisation, as monochromatic waves from different depths into the jet would under go different amounts of Faraday rotation, meaning the angles would not be coherent. This result is similar to that shown in \cite{eilek02}, which presents a RM map of binary HT galaxies in Abell 400, in which a jet from each HT galaxy crosses over the other. From the RM maps, it is possible to see structure preserved between the jets seamlessly, suggesting the RM structure is not from jets, which should have independent RM structure, but rather from a foreground Faraday screen. Such a screen could either be from either the Milky Way or the cluster in which the sources are embedded. PKS J0334-3900 is far above the Galactic plane, with a latitude of $-54.3^{\circ}$, so the magnetic field of the Milky Way is not expected to make a significant contribution to the RM. Estimations of the Galactic RM in the foreground of the cluster give values of 3.25 and 3.62 $\pm$ 5.65 rad m$^2$, from \cite{johnston-hollitt04} and \cite{oppermann11} respectively, leading us to conclude that the observed RM is due to the magneto-ionic plasma in Abell 3135. We find the RM structure along each jet is coherent at a scale of order $15^{\prime\prime}$ which is two times the beam size of $7.1^{\prime\prime} \times 7.1^{\prime\prime}$, as shown in Figure \ref{fig:rotationmeasure}. The average and standard deviation of RM\textsubscript{Synthesis} for the jets can be found in Table \ref{tab:RMstats}.  

An RM was calculated from the vector average linear polarisation of the jets. To do this, the jets were unresolved by using a beam size of 250 arcseconds. Then the Stokes Q and U of each jet was measured to calculate the vector averaged $P$ for each jet. From the two jets, the rotation measure was calculated using RM synthesis, using the same 12 intervals. The resultant RM was found to be 27.4 \& 5.5 rad m$^{-2}$ from the bottom and top jets respectively. Table \ref{tab:RMstats} compares these RMs, and the average RMs from the RM synthesis map. There is a difference between the P-vector average RM and the simple pixel based statistical analysis. In this case it was found that the P-vector average RM was dominated by a small number of high polarisation pixels. It is therefore more correct here to use the pixel based values to determine the average RM over each jet e.g. \cite{guidetti08} \& \cite{guidetti10}. 

\begin{table}
\caption{ Table comparing the RM values found in the top and bottom jets, from statistics applied to the pixels, and the vector average polarisation. $\mu$ and $\sigma$ are the mean and standard variation for the rotation measure derived using the pixel based analysis, assuming Gaussian statistics.}\label{tab:RMstats}
\begin{tabular}{l l l l}
\hline
Jet&$\mu$&$\sigma$&P-vector Average RM\\
\hline
&rad m$^{-2}$&rad m$^{-2}$&rad m$^{-2}$\\
\hline
Top&1.16&29.05&5.5\\
Bottom&7.44&24.88&27.4\\
Both&5.54&26.37\\\hline
\end{tabular}

\end{table}

\section{Discussion}

\subsection{Morphology of the Head-Tail galaxy}\label{sec:morphology}
PKS J0334-3900 has a morphology consisting of two bent jets and asymmetric hooks, which can be seen in Figure \ref{fig:cont1384}. Since both jets have a similar hook shape, the hooks  were likely created by a mechanism which affects each jet simultaneously. If the creation of the hooks was solely due to the motion of the ICM, the ICM would affect the jets independently and asymmetrically.  A simpler explanation for how the hooks were generated can be found by considering a possible orbital motion of the AGN and its companion (see Figure \ref{fig:StartOfJet}). This case would intuitively give a solution, which generates the similar scale and asymmetry between the hooks. The presence of close companions to HTs has been confirmed spectroscopically in the literature \citep{rose82,mao09}, and it has been suggested that a close companion is a necessary, but not sufficient condition to generate their morphology. If the possible orbital system here was falling into the cluster, or there were large-scale flows in the ICM due to a cluster merger, the HT galaxy would be affected as if there was a `wind' in the ICM. This wind would cause both jets to bend as a whole in the same direction, preserving the hook shapes generated from orbital motion.

\subsection{Modelling of the tailed radio galaxy}\label{sec:model}
\begin{figure*}
\vspace{-0.15cm}
\centering
\vbox
{\includegraphics[width=6.35in]{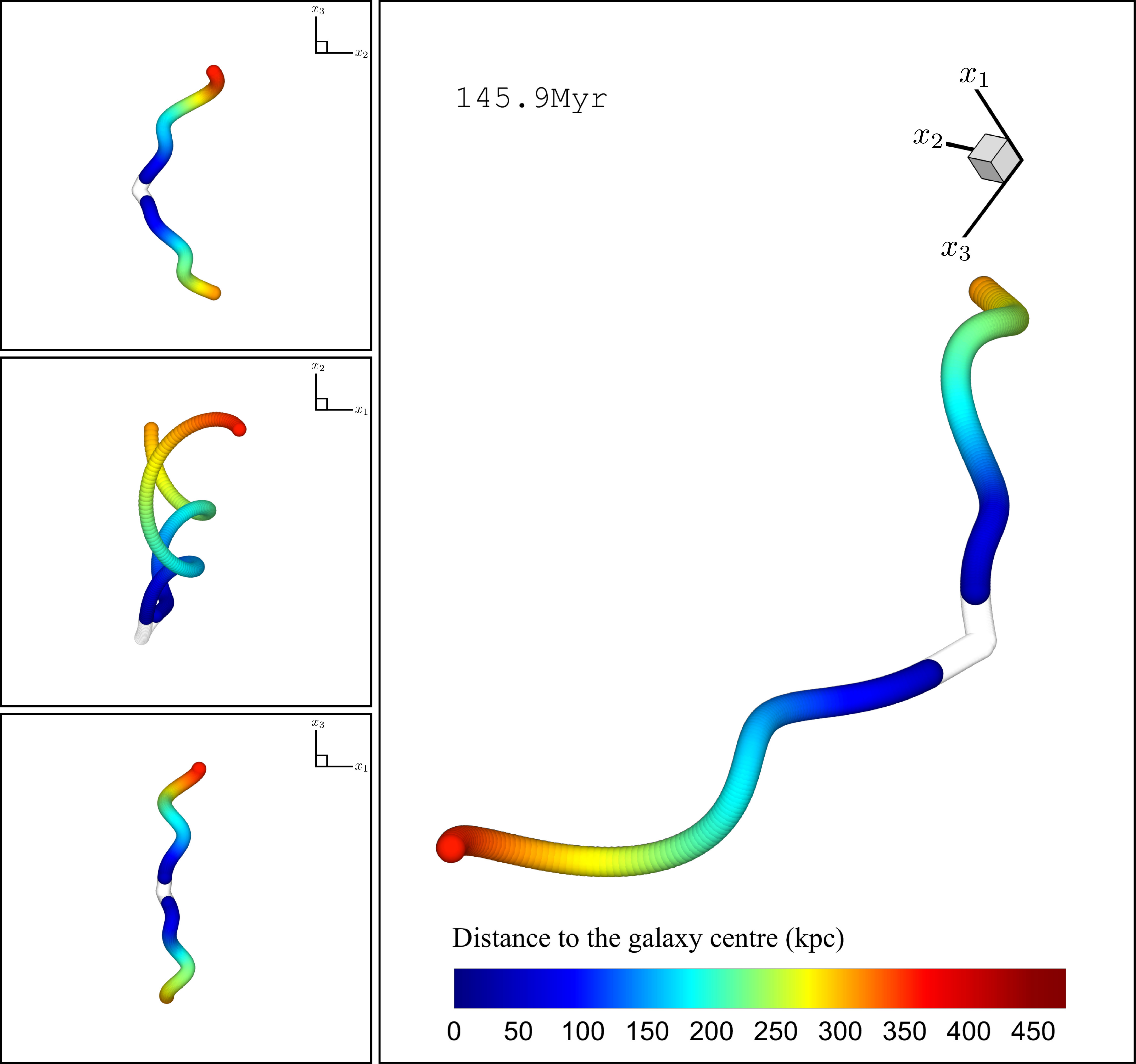}}
{\includegraphics[width=3.17in]{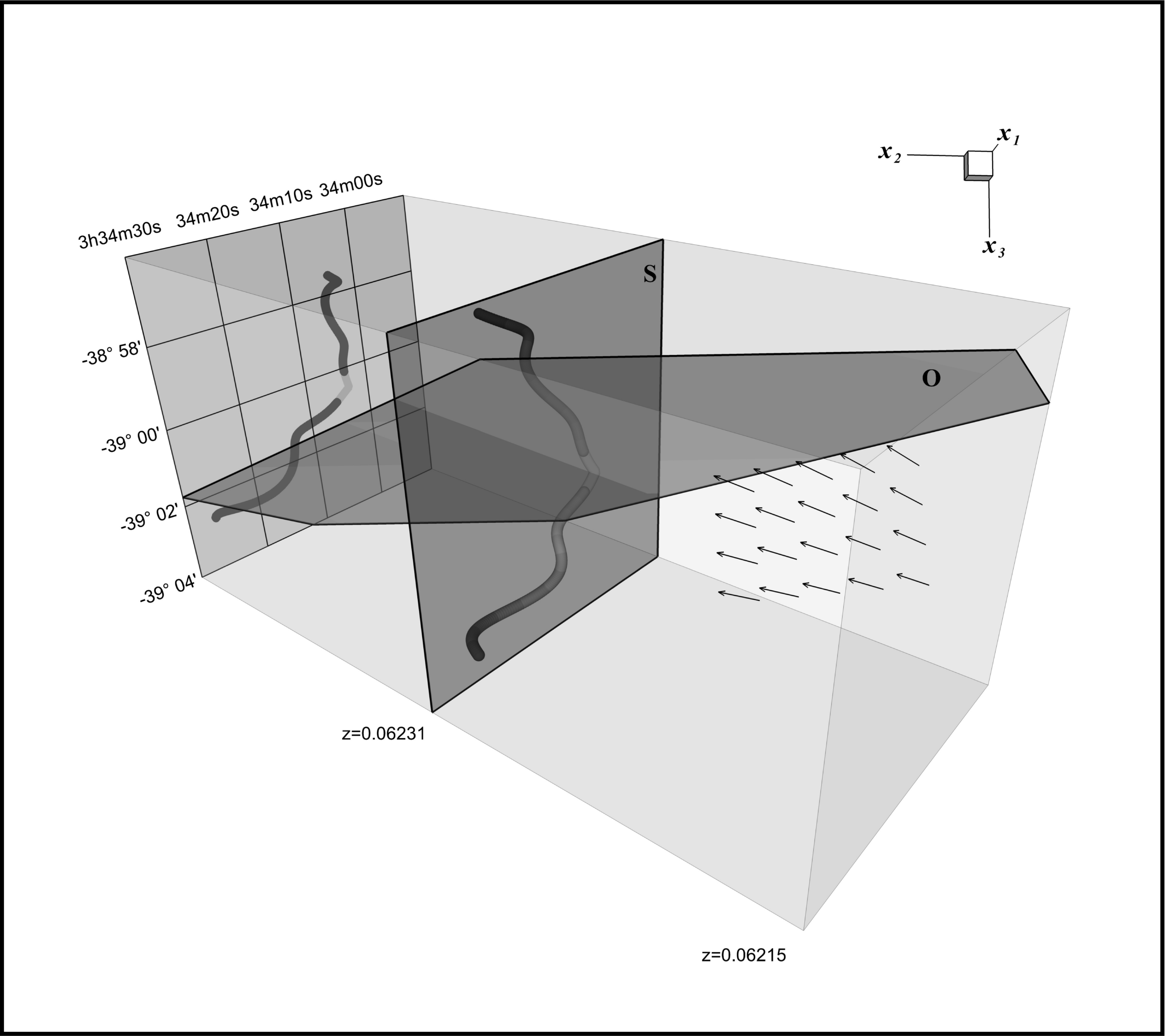}}
{\includegraphics[width=3.15in]{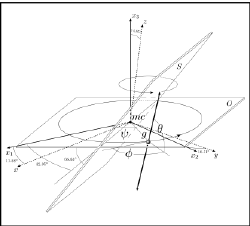}}
\vspace{-0.2cm}
\caption{{\bf Top figure:} Multi-view orthographic projections of the PKS-J0334-3900 model. The transparent region corresponds to the gap in the radio image close to the AGN. The top/bottom jets are moving away/towards us respectively. {\bf Lower left figure:} 3D orientations of the orbital plane $(O)$ and plane of sky $(S)$. Arrows in the top right of the box show the direction of the presumptive wind. {\bf Lower right figure:} Relative orientation of the coordinate system used in the model $(x_{1}x_{2}x_{3})$, and the equatorial coordinate system ($xyz$). The orbital plane lies on the $x_1-x_2$ plane. $mc$ and $g$ represent the mass centre of the system and the radio galaxy respectively.}\label{fig:diagram}
\end{figure*}

A simple simulation was performed to test if the hook and bent-tail morphology can be induced via orbital motion and a cluster wind. This model consists of a spinning and rotating source emitting a steady stream of spheres into two opposite directions. We establish a coordinate system, $x_1, x_2, x_3$ in which the model is centred on the AGN-companion mass centre, and defined such that the orbital plane lies on the $x_1-x_2$ plane, and $\psi_0=1.67$ rad sets the initial azimuthal position of the galaxy relative to $x_2$ axis (see lower right side of Figure \ref{fig:diagram}). Our model consists of symmetric and, to a first approximation, ballistic jets moving at constant speed. Ballistic jets have been frequently used as a first approximation  for radio galaxies \citep{blandford78, borne93, clarke96}. On kpc scales, the symmetric structure and luminosity of the jets in FRI sources suggest that the jets quickly decelerate to non-relativistic speeds \citep{laing02}, this is supported by both the radio emission on large-scales and the X-ray analysis.

The equations of motion of the spheres with the production rate of $R$ can be written as
\[
x_{lk} = (v^j_{lk}+v^o_{lk}+v^w_{lk})(t-\frac{l}{R})+x^o_{lk}
\]
for $k = 1, 2,$ and $ 3$, where $v^j_{lk}$, $v^o_{lk}$, and $v^w_{lk}$ are initial velocity components of the $l^{th}$ sphere due to jets, orbital motion, and wind respectively. These velocity components are defined by the following equations:
\[
v^j_{lk} =
\begin{cases}
v^j\cos(\theta)\cos(\phi) & k=1 \\
v^j\cos(\theta)\sin(\phi) & k=2 \\
v^j\sin(\theta) & k=3 \\
\end{cases}
\,\,,
\]
\[
v^o_{lk} =
\begin{cases}
-\psi^{\prime}r\sin(\psi) & k=1 \\
\psi^{\prime}r\cos(\psi) & k=2 \\
0 & k=3 \\
\end{cases}
\,\,,\,\,\,\,\,\, and \,\,\,\,\,\,\,\, v^w_{lk} = Constant,
\]
where $\theta$, $\phi$, and $\psi$ are the polar and azimuthal angles of the precession axis, and the azimuthal angle of the AGN in the orbit respectively. $x^o_{lk}$ represents the initial position of the AGN in the orbit.

In this simulation we assumed the jet's velocity to be $\sim$1 per cent of the speed of the light yielding $\gamma \sim$1, so the relativistic correction is not important. The model does not include energy loss or particle interactions. In \cite{blandford78}, a similar model was applied to the radio galaxy 3C 31, to test if the hook shapes in 3C 31 were due to orbital motion with a companion. However, the model of 3C 31 does not consider an interaction with the ICM, such as a `wind', or consider the possibility of the radio galaxy undergoing precession. The model of PKS J0334-3900 includes such variables required to generate the hook and bent tail morphology.

Parameters for the model were chosen using the available radio, optical and X-ray data. The radio structure and polarization vectors of the extended radio source were used to match the morphology, and the visible luminosity of the sources were used to calculate their mass. The model's initial conditions and parameters were determined by ensuring the best possible fit to the mean width of the radio jets. These parameters are given in Table \ref{tab:parameters}, and the model is displayed in Figure \ref{fig:diagram} (top). 

\begin{table}
\caption{Orbital and characteristic parameters for the PKS J0334-3900 model.}\label{tab:parameters}
\begin{tabular}{l l}
\hline
Observed Parameter & Value \\ 
\hline
Red magnitude of PKS J0334-3900 & 14.92 $\pm$ 0.06\\
Red magnitude of the companion & 15.92 $\pm$ 0.08 \\
Projected separation & 20.54 kpc\\
\hline
Dynamical Parameter & Value \\ 
\hline
J0334-3900 mass & $5.23^{+2.85}_{-2.00} \times 10^{12} M_\odot$ \\
The companion  mass & $1.44^{+0.42}_{-0.37} \times 10^{12} M_\odot$ \\
Actual separation & 35.21 kpc\\
Orbital velocity & 194.88 km/s\\
$x_{1}$ component of radial velocity & 516.54 km/s\\
$x_{2}$ component of radial velocity & 1162.21 km/s\\
$x_{3}$ component of radial velocity & -86.09 km/s\\
Jet velocity & 2060.98 km/s\\
Counterjet velocity & 1613.53 km/s\\
Eccentricity & 0\\
Initial orbital phase & $95.73^{\circ}$\\
obliquity & $8.10^{\circ}$\\
Precession period & 84.0 Myr\\
The dihedral angle & $65.63^{\circ}$\\
Evolution time & 145.9 Myr\\
\label{tab:param}
\end{tabular}
\end{table}

The major wide angle of PKS J0334-3900 emerges from its expanding tails' interacting with the surrounding ICM, or alternatively, it can be related to large-scale flows in the ICM. The relative motion of the HT through the intra-cluster medium generates strong ram pressure on the tails, giving the required angle between the jets. 

The difference in scale between the top and bottom jet cannot be explained by viewing angle alone. The ratio of the hook length to that of the total jet length is the same in each, despite the jets having different total lengths.  This means the foreshortening of the top jet seen in the image cannot be produced by rotation of the viewing angle in the simulation solely. Furthermore, this can be explained if there is a difference in the actual length of the jets. This can be described by variance in the jet and counter-jet velocity, or the ICM's density gradient; the top jet encounters a denser region. The latter is consistent with the overlay of the X-ray image shown in Section \ref{chandra}, which shows the end of the foreground (bottom) jet lies outside the X-ray contours, which is not the case for the top jet.

The minor features of the tail in the simulation are generated by the close encounter of two galaxies; the host galaxy and the smaller companion galaxy (see Figure \ref{fig:StartOfJet}) orbit a common centre of mass. Mass estimation of the system was made following the assumption that there is a power law correspondence between the halo mass and the red luminosity of the galaxy:
\[
M=M_{\star}\left(\frac {L}{L_{\star}}\right)^{\beta}
\]
with $M_{\star}=10.73 \pm 2.53 (10^{11}h^{-1}M_\odot)$, $L_{\star}=1.51 \pm 0.04 (10^{10}h^{-2}L_\odot)$, and $\beta=1.4 \pm0.2$, optimized for the early type galaxies \citep{guzik02}.  Unfortunately the r-band rest frame luminosity of the galaxies could not be calculated via fitting of a template spectral energy distribution (SED) due to the lack of the other luminosity bands. However, examination of k-corrections for other r-band magnitudes at a redshift of 0.06 demonstrates at worst a maximum correction of 0.1 would be likely \citep{blanton07}, while a k-correction based on an empirical polynomial fit \citep{chili10,westra10} would result in something more like 0.08. As the uncertainty in the mass estimate is dominated by other factors, the application of even the largest likely k-correction at this redshift gives a mass which is almost indistinguishable from a mass derived from an un-corrected r-band magnitude. Without a proper SED fit, these other methods introduce other assumptions, which may not hold. For this reasons, non rest frame red magnitudes were used for the mass calculations which is given in Table \ref{tab:param}. These estimations give rise to additional free parameters, such as the dark matter portion of the general mass in the galaxies and also dynamical attributes such as eccentricity of the orbit. Nevertheless, uncertainty in these estimations would not create major issues since this is a qualitative way to understand the responsible mechanisms forming the morphology. 

It is essential to emphasize that the proposed model is a qualitative way to explain the morphology of this HT. In other words, due to a number of free parameters, which were not possible to determine from the available radio and optical data, this model is merely one possible way to justify the morphology.

\subsection{HT galaxies as probes of the intra-cluster magnetic field}
Using the model in Section \ref{sec:model}, we find the angle between the jets along the line of sight is 120 $\pm 5 ^{\circ}$, the linear length is 346 $\pm$ 32 kpc in the bottom jet and 170 $\pm$ 16 kpc in the top jet. So we find the distance between the jets, at halfway down them, is 154 $\pm$ 16 kpc. Using an average value of electron density, $1.09 \times 10^{-3}$ cm$^{-3}$ derived in Section \ref{chandra}, and the difference in RM between the jets, we find an average intra-cluster magnetic field, along the line of sight, to be 0.05 $\pm 0.02 \mu$G. Supposing that $\langle B\rangle = \sqrt{3}\langle B_{\parallel}\rangle$ \citep{eilek02}, we find the average magnetic field to be 0.09 $\pm 0.03 \mu$G. However, this is for the simplest case of a uniform magnetic field within the cluster, for more complicated fields it is expected to be larger due to random rotations in the cluster. \cite{eilek02} shows how it would be possible to compare the magnitude of $B$ for different models of magnetic fields causing the Faraday rotation, however, that is beyond the scope of this paper. 

\section{Conclusion}
In this work we present new radio, optical and X-ray analysis for A3135 and the associated HT galaxy PKS J0334-3900. We present a new cluster mean redshift of 0.06228 $\pm$ 0.00015, which is lower than the previously published value due to the presence of a heretofore undetected subgroup, with 0.066 $ \leq z <$ 0.070. We present new Chandra images of the clusters which  reveal an extended X-ray morphology and provide evidence of a recent merger. Using new radio analysis at multiple frequencies we present a detailed analysis of the central HT galaxy PKS J0334-3900. We show that the morphology and use of linear polarisation of an HT galaxy can be used to determine not only the properties of the HT itself, but also as a probe of conditions in the ICM. The morphology and estimated age of the jets were used to make a simulation. This simulation showed that if the HT is falling into the cluster along with its companion about which it is undergoing an orbital motion, the observed morphology can be produced. Alternatively, large-scale flows in the ICM due to a cluster-cluster merger in combination with the orbital motion of PKS J0334-3900 and it's companion about their centre of mass can produce the same morphology. This simulation was used to derive a likely angle between the jets along the line of sight to be 125 degrees. Furthermore, we estimate the upper limit on the age of the jets as 146 Myr, which also provides a time frame where the direction of the `wind' resultant from infall is recorded in the morphology of the jets. 

Additionally, rotation measure maps for each jet showed that the magnetic field was ordered on scales of roughly 15$^{\prime\prime}$ (18 kpc), and the averaged RMs for each jet were 1.16 and 7.44 rad m$^{-2}$ for the top and bottom respectively. We attribute this difference to the different path lengths of light from each jet passing through the ICM. 
Using the linear lengths of the jets and the angle derived from the simulation, we obtained the likely distance between the jets along the line of sight. This distance was used in conjunction with the difference in average RM between the jets and average electron density, to calculate an average cluster magnetic field along the line of sight. This gave a lower limit of 0.05 $ \pm 0.02 \mu$G for the line of sight component which suggests a lower limit for the total cluster field of 0.09 $ \pm 0.03 \mu$G. While the magnetic field is slightly weaker than typical literature values for cluster fields using other methods, which show ranges between 0.1 to a few $\mu$G dependent on different locations in the cluster (see for example comparison of different values across A3667 \citep{mjh04b}), as our value is a lower limit, this does not present a problem. This is a novel method for estimating the cluster magnetic field which can be added to the repertoire of methods to probe cluster magnetic fields. 

In the next decade a number of radio telescopes are expected to be built or upgraded including; LOFAR \citep{rottgering03}, MWA \citep{tingay12}, ASKAP \citep{johnston07}, MeerKAT \citep{jonas09}, WRST and the EVLA. These instruments are anticipated to conduct a series of all-sky continuum radio surveys such as EMU \citep{norris11} and WODAN \citep{rottgering11} and undertake a wide range of survey science projects \citep{norris13}. These next generation surveys will produce catalogues of radio galaxies numbering in the tens of millions \citep{norris11}, within such surveys we expect to find between tens to hundreds of thousands of Head-Tailed galaxies \citep{mao11,Dehghan11a}, which will form an important population for cluster physics. In recent years, the use of HTs as probes of local over densities in their near by environment has increased somewhat. This includes as signposts for detecting clusters \citep{blanton03,smolcic07} and local over densities \citep{mao10}. However, despite this oft cited use, few studies have fully utilized HTs as environmental probes. This work highlights how rich HT galaxies will be for understanding clusters in next generation radio surveys.

\section{Acknowledgements}
We thank Ron Ekers and Robert Laing for valuable discussions on rotation measure statistics and Leith Godfrey for discussions on jet mechanics. LP is grateful for support via Victoria University of Wellington Faculty of Science and Marsden Development Fund research grants awarded to MJ-H. SD is supported by a Victoria University of Wellington Vice-Chancellor's Strategic Scholarship in support of the research of M.J-H.
The Australia Telescope Compact Array telescope is part of the Australia Telescope which is funded by the Commonwealth of Australia for operation as a National Facility managed by CSIRO. The Digitized Sky Survey was produced at the Space Telescope Science Institute under US Government grant NAG W-2166 and is based on photographic data obtained using The UK Schmidt Telescope. The UK Schmidt Telescope was operated by the Royal Observatory Edinburgh, with funding from the UK Science and Engineering Research Council, until 1988 June, and thereafter by the Anglo-Australian Observatory.
Original plate material is copyright (c) of the Royal Observatory Edinburgh and the Anglo-Australian Observatory.
The plates were processed into the present compressed digital form with their permission. 
SuperCOSMOS Sky Survey material is based on photographic data originating from the UK, Palomar and ESO Schmidt telescopes and is provided by the Wide-Field Astronomy Unit, Institute for Astronomy, University of Edinburgh.
This research has made use of the NASA/IPAC Extragalactic Database (NED) which is operated by the Jet Propulsion Laboratory, California Institute of Technology, under contract with the National Aeronautics and Space Administration. This research has also made use of NASA's Astrophysics Data System.
\bibliographystyle{mn}

\bibliography{Abell3135}{}

\label{lastpage}
\end{document}